\documentclass[aps,prb,10pt,superscriptaddress,notitlepage,twocolumn]{revtex4-1}
\usepackage{amsmath}

\usepackage{graphicx}
\graphicspath{{figures/}{./}}

\usepackage[colorlinks=true,urlcolor=blue,linkcolor=blue,citecolor=red]{hyperref}

\usepackage{xcolor}
\usepackage{tikz}
\begin{document}

\title{Optical characterization of multilayer
\texorpdfstring{$\alpha\textrm{-In}_{2}\textrm{Se}_{3}$}{alpha-In2Se3}}

\author{Yujin Cho}%\email{ycho@physics.utexas.edu}
    \affiliation{Department of Physics, University of Texas at Austin,
                 Austin, Texas 78712, USA}
\author{Sean M. Anderson}\email{sma@cio.mx}
    \affiliation{Centro de Investigaciones en \'Optica, 
                Le\'on, Guanajuato 36000, M\'exico}
\author{Bernardo S. Mendoza}%\email{bms@cio.mx}
    \affiliation{Centro de Investigaciones en \'Optica, 
                Le\'on, Guanajuato 36000, M\'exico}
\author{Shun Okano}%\email{shun.okano@s2015.tu-chemnitz.de}
    \affiliation{Semiconductor Physics, Chemnitz University of Technology,
                09107, Chemnitz, Germany}
\author{N. Arzate}%\email{narzate@cio.mx}
    \affiliation{Centro de Investigaciones en \'Optica, 
                Le\'on, Guanajuato 36000, M\'exico}
\author{Anatoli I. Shkrebtii}%\email{anatoli.chkrebtii@uoit.ca}
    \affiliation{Ontario Tech University,
                 Oshawa, ON, L1G 0C5, Canada}
\author{Di Wu}%\email{diwu@utexas.edu}
    \affiliation{Department of Physics, University of Texas at Austin,
                 Austin, Texas 78712, USA}
\author{Keji Lai}%\email{kejilai@physics.utexas.edu}
    \affiliation{Department of Physics, University of Texas at Austin,
                 Austin, Texas 78712, USA}
\author{Ram\'on Carriles }%\email{ramon@cio.mx}
    \affiliation{Centro de Investigaciones en \'Optica, 
                Le\'on, Guanajuato 36000, M\'exico}
\author{D. R. T. Zahn}%\email{zahn@physik.tu-chemnitz.de}
    \affiliation{Semiconductor Physics, Chemnitz University of Technology,
                09107, Chemnitz, Germany}
\author{M. C. Downer}%\email{downer@physics.utexas.edu}
    \affiliation{Department of Physics, University of Texas at Austin,
                 Austin, Texas 78712, USA}
\date{\today}

\begin{abstract}
Ferroelectric materials possess spontaneous electric polarization below the phase
transition temperatures, which are switchable with an external electric field. 2D
ferroelectric materials have many potential uses, but an understanding of how
this spontaneous ferroelectricity changes with different physical
properties is crucial to properly engineer these materials for future
applications. These properties can be effectively probed using optical
techniques, which is excellent motivation for carrying out a systematic study of
various opto-electronic properties using spectroscopic techniques. In
particular, this work focuses on the evolution of the linear and nonlinear
optical responses of layered and bulk $\alpha\textrm{-In}_{2}\textrm{Se}_{3}$
for different nano-flake thicknesses and orientations, using
high-resolution spectroscopic measurements and \emph{ab initio} 
density functional theory (DFT) and time-domain DFT (TDDFT)
calculations.

%Our results for various quintuple layers (QLs) of
%$\alpha\textrm{-In}_{2}\textrm{Se}_{3}$ indicate that rotational anisotropic
%second harmonic generation (SHG) is sensitive enough to discern individual QL
%orientation, while 
Nonlinear second-harmonic generation (SHG) spectroscopy measurements feature a broad resonant peak
centered around 1.4\,eV. The intensity of the SHG spectra increases with the
number of layers up to three quintuple layers, and then steadily decreases for
larger numbers. With support from DFT calculations, we found that the net 
ferroelectric polarization on 2 QLs is most likely zero, while thicker QLs have 
non-zero polarization.
We also present transmission measurements for layered samples
over a photon energy range of 1.5--4\,eV, along with ellipsometry data for the
complex index of refraction for bulk $\alpha\textrm{-In}_{2}\textrm{Se}_{3}$;
these are all compared with \emph{ab initio} DFT and TDDFT calculations. The
linear response is not as sensitive to structural variations as the SHG
spectra, but allow us to discern critical point transitions and to extract the
optical band gap values, that are also corroborated with $G_{0}W_{0}$
calculations. The thorough study of the linear and nonlinear optical properties of 
$\alpha\textrm{-In}_{2}\textrm{Se}_{3}$ will be useful for potential 
applications as optoelectronic devices.  
\end{abstract}

% \pacs{78.67.Pt,78.68.+m,71.15.Mb}
\maketitle

%%%%%%%%%%%%%%%%%%%%%%%%%%%%%%%%%%%%%%%%%%%%%%%%%%%%%%%%%%%%%%%%%%%%%%%%%%%%%%%%
%%%%%%%%%%%%%%%%%%%%%%%%%%%%%%%%%%%%%%%%%%%%%%%%%%%%%%%%%%%%%%%%%%%%%%%%%%%%%%%%

\section{Introduction}\label{sec:intro}

III-VI semiconductor materials exhibit a variety of structural and electronic
properties, due mostly to the complex nature of the valence electrons in the
group VI elements. These compound materials demonstrate a rich variety of
compositions. Among numerous bulk III-VI crystals, recent findings show that
these materials also exist in the form of two-dimensional (2D) hexagonal
nanofilms that are only a few atoms thick \cite{boukhvalovNM17, demirciPRB17,
linCEC18}. In particular, III$_{2}$-VI$_{3}$ compounds (also known as
A$_{2}$B$_{3}$-type chalcogenides) composed of indium (In) and selenium (Se) can
form several single-phase and layered In-Se compounds; namely, InSe
\cite{henckPRM19}, $\textrm{In}_{2}\textrm{Se}_{3}$,
$\textrm{In}_{3}\textrm{Se}_{4}$ \cite{rhyeeN09, hanCEC14} and
$\textrm{In}_{4}\textrm{Se}_{3}$, all of which have been obtained experimentally
\cite{debbichiAP14, bandurinNN16, liuJACE18}.

The switchability of ferroelectric polarization below the transition temperature
\cite{dingNC17} is desirable for many applications
\cite{scottSCI07}, and particularly useful for storing information in memory
devices at higher speeds with less power consumption \cite{millerJAP92,
parkNat99, wanNS18}. In addition, a growing interest in device miniaturization
naturally leads to search for new thin film ferroelectrics \cite{ponathNC15,
choAPL18}. However, when the thickness shrinks beyond some critical value,
usually on the order of a few nanometers, strong depolarization fields tend to
suppress the ferroelectricity; this has been a significant challenge for
traditional perovskite thin films \cite{fongSCI04, meyerPRB01}. The
strain effect induced at the interface between the substrate and the film
affects the spontaneous polarization near the interface \cite{hanNC14,
dawberRMP05}. Fortunately, 2D ferroelectrics can potentially overcome these
difficulties. Lack of out-of-plane chemical bonding of 2D materials reduces
misfit strain at the interface between the substrate and the films
\cite{changSCI16} and the spontaneous polarization remains switchable at a few
layers. In-plane ferroelectricity of 2D materials has been demonstrated in
several materials such as MoS$_{2}$, SnTe, and Phosphorene analogues
\cite{shirodkarPRL14, changSCI16, wuNL16}.

Layered $\textrm{In}_{2}\textrm{Se}_{3}$ is a 2D van der Waals (vdW), nonplanar,
and quasi-two-dimensional semiconducting material \cite{liPRB18} that has
recently garnered significant interest with a large variety of identified
applications \cite{boukhvalovNM17, demirciPRB17, linCEC18, wanAFM19, yangAMT19}.
It has several advantageous optoelectronic properties; for instance,
$\alpha\textrm{-In}_{2}\textrm{Se}_{3}$ demonstrates a thickness-dependent band
gap \cite{queredaAOM16} which enables tuning of its dielectric function, and
thus, optical properties \cite{wuNL15, huRA17}. It is also temperature tunable:
non-centrosymmetric $\alpha\textrm{-In}_{2}\textrm{Se}_{3}$ transforms into
centrosymmetric $\beta\textrm{-In}_{2}\textrm{Se}_{3}$ at 475\,K \cite{taoNL13},
which is propitious for phase-change memory applications \cite{yuAPL07,
choiAM17}.

In contrast to the single-atomic graphene-like materials, or the more complex
phosphorene and MoS$_{2}$, the basic $\textrm{In}_{2}\textrm{Se}_{3}$ layer
consists of five alternating Se and In atoms that form the so-called quintuple
layer (QL). Numerous structural variations of the QLs are possible
\cite{dingNC17}, with the two most energetically stable layered structures
belong to the $\alpha$ and $\beta$ phases. Both are semiconducting with an
energy bandgap in the optical range, with widely varying reported experimental
and theoretical bandgap values. The $\alpha$ phase has the lowest total energy
and belongs to the $R3m$ space group; in contrast, the $\beta$ phase is less
energetically stable and belongs to the $R\bar{3}m$ space group.
% It actually consists of two energetically degenerate variants: one
% zincblende-like (FE-ZB$^{\prime}$) and one wurtzite-like (FE-WZ$^{\prime}$)
% \cite{dingNC17}. As we will show below, both FE-ZB$^{\prime}$ and
% FE-WZ$^{\prime}$ structures produce very similar band structures and optical
% responses.
Recent 
% but has recently been demonstrated in CuInP$_{2}$S$_{6}$ down to thicknesses
% of 4\,nm \cite{liuNC16}.
experimental work has demonstrated that $\alpha\textrm{-In}_{2}\textrm{Se}_{3}$
possesses room-temperature out-of-plane ferroelectricity down to thicknesses of
a few layers \cite{zhouNL17, wanNS18, xiaoPRL18, xueAFM18, zhengSA18}, and
expected to persist even down to a single QL \cite{dingNC17}. However, the
presence of a strong dipole moment favors the formation of 2D domains with
opposite dipole orientations which reduces the electrostatic energy; for
instance, the potential barrier for changing the dipole orientation is around
0.07 eV \cite{dingNC17}. Experimental results \cite{wanNS18} indicate that a few
volts of applied electric potential in the perpendicular direction can switch
the dipole directions of the QLs. Moreover, the magnitude of the spontaneous
polarization is not linear to the thickness; it is expected to be maximized at
three QLs, and decreases with additional layers \cite{dingNC17}. Multi-QL
systems have weak-vdW bonded layers, and so we must consider several QL
orientations with parallel and opposite dipole moments \cite{wuNL15, zhouNL17,
wanNS18}.
Since the experimentally observed symmetry for single and multiple QLs at room
temperature is consistent with the $R3m$ space group \cite{wuNL15, wanNS18},
% Therefore, for the theoretical calculations presented in this work, we will
% only consider the $\alpha$ phase. with both FE-ZB$^{\prime}$ and
% FE-WZ$^{\prime}$ QLs taken into consideration.
we will focus exclusively on the optical characterization of 
$\alpha\textrm{-In}_{2}\textrm{Se}_{3}$ as a promising ferroelectric material.

We can gain a critical understanding of this material from its optical properties;
for instance, the transmission and absorption spectra are directly related to
different intrinsic quantities such as the dielectric function and the
electronic band structure. Likewise, optical second-harmonic generation (SHG) is
an efficient and non-destructive spectroscopic method, that is very sensitive to
even small changes in the atomic structure and symmetry properties (including
centrosymmetry). SHG could allow us to characterize the polarization dependence
of the optical response, and potentially elucidate information about the
microscopic structure of the individual QLs. These quantities can be explored
experimentally using spectroscopic methods, and theoretically determined from
\emph{ab initio} calculations. Knowledge of how the spontaneous polarization
changes with the thickness and microscopic arrangement of this material is
crucial for engineering $\alpha\textrm{-In}_{2}\textrm{Se}_{3}$ towards device
applications; therefore, this is excellent motivation for carrying out a
systematic study of various opto-electronic properties of
$\alpha\textrm{-In}_{2}\textrm{Se}_{3}$, using several experimental and
theoretical spectroscopic techniques. In particular, this work focuses on the evolution of the linear and nonlinear optical responses of layered and bulk $\alpha\textrm{-In}_{2}\textrm{Se}_{3}$ for different nano-flake thicknesses and
orientations. The results presented from high-resolution spectroscopic measurements are complemented by \emph{ab initio} calculations that allow insight into the intrinsic opto-electronic properties of the material, to elucidate the ferroelectric nature and microscopic characteristics.
% We present a complete set of optical transmission spectra and dielectric
% functions determined from spectroscopic ellipsometry (SE) measurements; we
% also carry out \emph{ab initio} calculations within the time-dependent density
% functional theory (TDDFT) and density functional theory within local-density
% approximation (DFT-LDA) frameworks. The dielectric function is sensitive to
% small changes in the electronic band structure; thus, we can elucidate the
% changes in the electronic properties as a function of the material thickness.
% Previous works have provided some experimental evidence for the
% thickness-dependence of the dielectric functions at 1 GHz \cite{wuNL15}, and
% the band gap value for flake thicknesses ranging from 3.1 up to 25\,nm
% \cite{queredaAOM16}. However, a thorough and systematic analysis of the
% dielectric function (and the associated transmission and absorption spectra)
% in the visible light range, and of the optical band gaps from monolayer to
% bulk, has not yet been reported.

This paper is organized as follows. In section \ref{sec:methods}, we present the
experimental and computational methods used to carry out the study of the
optical properties of $\alpha\textrm{-In}_{2}\textrm{Se}_{3}$. In section
\ref{sec:results}, we present our measured and calculated spectra for both the
layered and the bulk material, obtained from direct spectroscopic measurements
and \emph{ab initio} calculations. Lastly, we list our conclusions and
final remarks in section \ref{sec:conclusion}.

%%%%%%%%%%%%%%%%%%%%%%%%%%%%%%%%%%%%%%%%%%%%%%%%%%%%%%%%%%%%%%%%%%%%%%%%%%%%%%%%
%%%%%%%%%%%%%%%%%%%%%%%%%%%%%%%%%%%%%%%%%%%%%%%%%%%%%%%%%%%%%%%%%%%%%%%%%%%%%%%%

\section{Methods}\label{sec:methods}

%%%%%%%%%%%%%%%%%%%%%%%%%%%%%%%%%%%%%%%%%%%%%%%%%%%%%%%%%%%%%%%%%%%%%%%%%%%%%%%%
%%%%%%%%%%%%%%%%%%%%%%%%%%%%%%%%%%%%%%%%%%%%%%%%%%%%%%%%%%%%%%%%%%%%%%%%%%%%%%%%

\subsection{Experimental Methods}

The $\textrm{In}_{2}\textrm{Se}_{3}$ nano-flakes were grown on fluorophlogopite
mica substrates by vapor phase deposition (VPD), which provides highly
crystalline samples and well-controlled thickness profiles \cite{linJACS13,
wuNL15, zhengNC15, zhouNL17}. The flakes were synthesized by vdW epitaxy in a
1--inch diameter horizontal quartz tube furnace (Lindberg Blue M HTF55667C). Bulk
$\textrm{In}_{2}\textrm{Se}_{3}$ powder ($99.99\,\%$, Alfa Aesar) was placed at
the center and heated to $740\,^{\circ}$C. The vapor flowed downstream with
30-200 SCCM argon gas at 20 Torr and deposited on mica substrates, placed
7--12\,cm away from the heated center, forming layered
$\textrm{In}_{2}\textrm{Se}_{3}$ flakes, typically a few tens of microns wide.
After 10 minutes, the tube was cooled down to room temperature at a rate slower
than $5\,^{\circ}$\,C/min. Like most 2D materials,
$\textrm{In}_{2}\textrm{Se}_{3}$ has a hexagonal structure that deposits in
triangular flakes, except for single quintuple layer flakes, which instead form
rounded shapes on the mica substrate (see Figures \ref{fig:micrographs}a and \ref{fig:micrographs}b). We
confirmed the thickness and the crystallinity of the layers using atomic force
microscopy and SHG microscopy, verifying the presence of the $\alpha$-phase of
$\textrm{In}_{2}\textrm{Se}_{3}$ with rotational anisotropic SHG (RASHG)
microscopy. This method has been previously used with success to synthesize
$\alpha\textrm{-In}_{2}\textrm{Se}_{3}$, having been confirmed by transmission
electron microscopy \cite{wuNL15}.

Optical transmission spectroscopy was performed on a micro-imaging ellipsometer
Nanofilm EP4 (Accurion GmbH), equipped with a Xenon arc lamp with a wavelength
range from 950--250\,nm (photon energy 1.3\,eV to 4.96\,eV, respectively), at
normal incidence. A single diffraction grating produced a monochromatic incident
light. Reflected light was collected and imaged with $\sim1$ $\mu$m spatial
resolution to an array with a Nikon 50$\times$ long working distance microscope
objective (N.A. = 0.45), enabling signals reflected from within the boundaries
of a single tens-of-micron-sized flake to be analyzed. For the SHG spectroscopy,
we used a Ti:Sapphire laser (Coherent Chameleon Vision II) operating at an
80\,MHz repetition rate with 140\,fs pulse duration. The wavelength of the laser
was tunable from 680\,nm to 1040\,nm. The laser light is incident on the sample
at 45$^{\circ}$ and the beam spot size was around 3 $\mu$m. We normalized the
measured SHG intensity from $\alpha\textrm{-In}_{2}\textrm{Se}_{3}$ to that of
$\alpha$-quartz to better estimate the magnitude of the
$\boldsymbol{\chi}^{(2)}$ tensor components. A $\lambda/2$-waveplate and a
linear polarizer were used to select incoming polarization and SHG polarization,
respectively. All measurements presented here were checked for consistency by
collecting data from several regions on the same sample, as well as from
different flakes with the same number of QLs.

%%%%%%%%%%%%%%%%%%%%%%%%%%%%%%%%%%%%%%%%%%%%%%%%%%%%%%%%%%%%%%%%%%%%%%%%%%%%%%%%
%%%%%%%%%%%%%%%%%%%%%%%%%%%%%%%%%%%%%%%%%%%%%%%%%%%%%%%%%%%%%%%%%%%%%%%%%%%%%%%%

\subsection{Computational methods}\label{sec:comp}

We relaxed the atomic positions for each QL variant using the repeated slab
approach with the Quantum Espresso \cite{giannozziJPCM17} software package,
within the DFT-LDA framework with plane-waves and pseudopotentials using a
60\,Ry cutoff. Single QLs of either structure belong to the $R3m$ space group
with a lattice constant of $a = 4.106$\,\r{A} \cite{dingNC17}. We modeled
several weakly vdW-bonded multi-QL configurations to determine the
optimal interlayer distances and QL stacking arrangements. In agreement with
previous calculations \cite{wuNL15, dingNC17, wanNS18}, we found that the most
energetically stable configuration is an ABC QL stacking \cite{dingNC17} with an
inter-QL separation of 2.95\,\r{A}. This separation, as well as the atomic
structure of the multi-QL systems, does not depend on the dipole orientation of
the individual QLs within the stack.

The calculation of the nonlinear susceptibility tensor
$\boldsymbol{\chi}(-2\omega;\omega,\omega)$ and the dielectric functions for
layered $\alpha\textrm{-In}_{2}\textrm{Se}_{3}$ were carried out with the TINIBA
code \cite{tiniba} within the independent-particle DFT-LDA framework
\cite{andersonPRB15}. First, we calculated the electronic wave-functions using
the ABINIT code \cite{gonzeCPS09, abinit} with a planewave basis set with an
energy cutoff between 20 to 40 Hartree, and Troullier-Martins LDA
pseudopotentials \cite{troullierPRB91}. The spectra were properly converged with
several thousand \textbf{k}-points in the IBZ. The contribution from the
nonlocal part of the pseudopotentials was carried out using the DP code
\cite{olevanoDP, reiningEXC} with a basis set of at least 5000 planewaves. The
total number of bands included in the calculation was at least $12\,(N + 1)$,
where N is the total number of QLs. We included a vacuum region between 30 to
130\,\r{A} (depending on the number of QLs) to compensate for the net dipole
moment present in some stacking arrangements and to avoid spurious wave-function
tunneling. All results presented below are thus normalized to the
vacuum region included in the total supercell height \cite{tancognePRB15};
yielding consistent results regardless of vacuum region size. The components of
$\boldsymbol{\chi}(-2\omega;\omega,\omega)$ are also properly normalized to
obtain the correct surface units of pm$^{2}$/V. Quasiparticle effects are
included, when pertinent, via a rigid scissors approach.

The calculation of the dielectric function for bulk
$\alpha\textrm{-In}_{2}\textrm{Se}_{3}$ was carried out within the TDDFT
framework using the DP \cite{olevanoDP, reiningEXC} code, where the
independent-particle response function was constructed using Kohn Sham (KS)
orbitals, within the random phase approximation (RPA) including local-field
effects. The electronic wave-functions were calculated using the ABINIT code
\cite{gonzeCPS09, abinit} and a planewave basis set with an energy cutoff of 15
Hartrees and Troullier-Martins LDA pseudopotentials \cite{troullierPRB91}, and
at least 1700 \textbf{k}-points in the irreducible Brillouin zone (IBZ) with 18
unoccupied states (30 total).

Lastly, we carried out $G_{0}W_{0}$ \cite{hedinPR65} calculations in order to
obtain self-corrected band gap values at the $\Gamma$ point. These were carried
out using the ABINIT code \cite{gonzeCPS09, abinit} using similar parameters to
those described above. We obtained values for 1 QL, 2 QLs, and bulk. These
values were used to adjust the required eigen-energies via a rigid scissors
approach for the different spectroscopic calculations above.

%%%%%%%%%%%%%%%%%%%%%%%%%%%%%%%%%%%%%%%%%%%%%%%%%%%%%%%%%%%%%%%%%%%%%%%%%%%%%%%%
%%%%%%%%%%%%%%%%%%%%%%%%%%%%%%%%%%%%%%%%%%%%%%%%%%%%%%%%%%%%%%%%%%%%%%%%%%%%%%%%

\section{Results}\label{sec:results}

%%%%%%%%%%%%%%%%%%%%%%%%%%%%%%%%%%%%%%%%%%%%%%%%%%%%%%%%%%%%%%%%%%%%%%%%%%%%%%%%
%%%%%%%%%%%%%%%%%%%%%%%%%%%%%%%%%%%%%%%%%%%%%%%%%%%%%%%%%%%%%%%%%%%%%%%%%%%%%%%%

\subsection{RASHG and SHG Spectroscopy}

\begin{figure}[t]
\centering
\includegraphics[width=\linewidth]{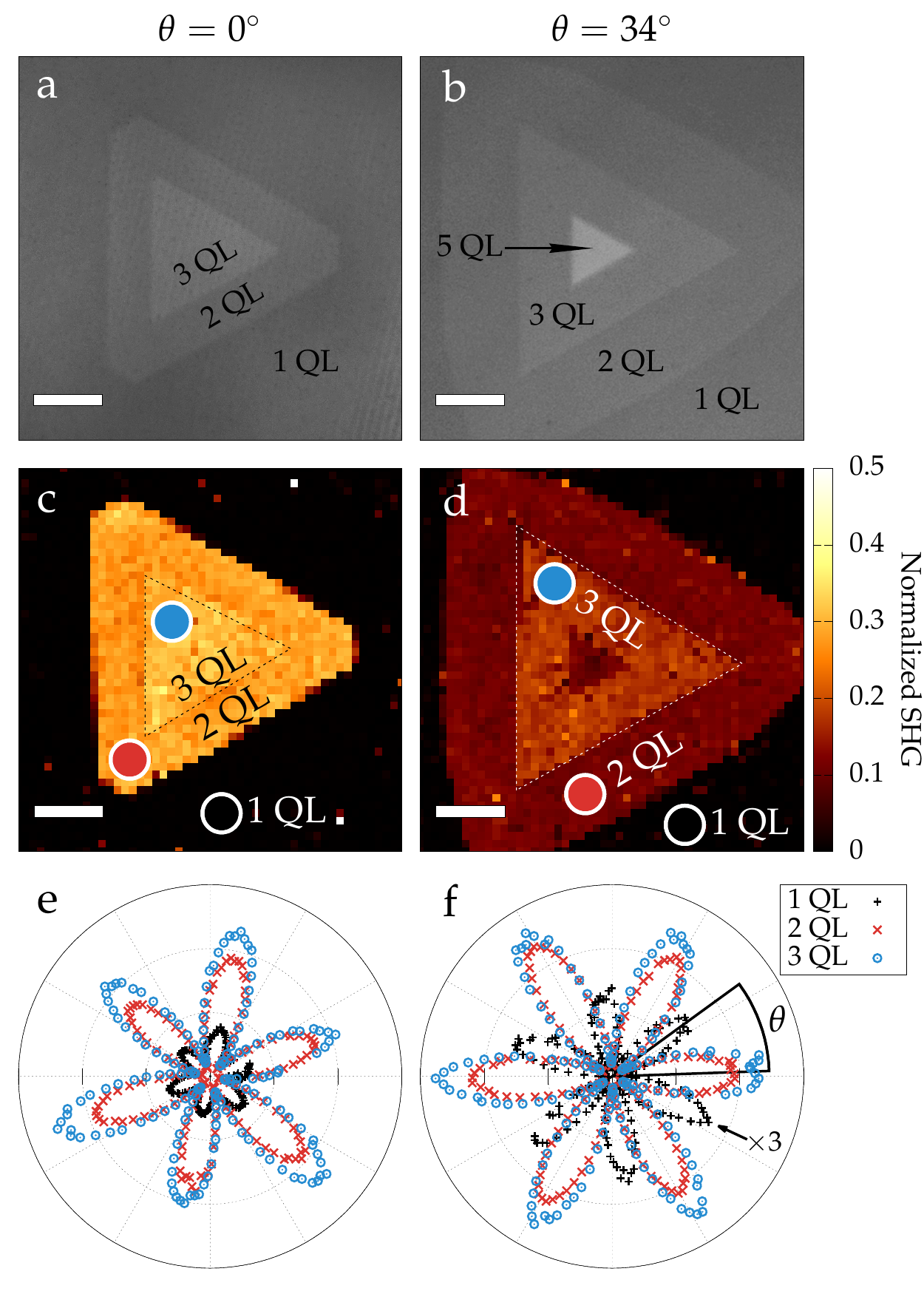}
\caption{\textbf{a}--\textbf{b}, optical images of two
$\alpha\textrm{-In}_{2}\textrm{Se}_{3}$ samples used in this work.
\textbf{c}--\textbf{d}, SHG micrographs of the same samples. The white bars
represent 50 $\mu$m.. The number of QLs are labeled in the appropriate regions.
\textbf{e}--\textbf{f}, RASHG patterns measured from the samples; line colors
correspond to the colored dots in \textbf{c} and \textbf{d}. $\theta$ is the
relative angle of crystallographic orientation between QLs. SHG measurements
(\textbf{c}--\textbf{f}) where taken at a fundamental wavelength of 780\,nm
(1.59\,eV).}
\label{fig:micrographs}
\end{figure}

Figures \ref{fig:micrographs}a and \ref{fig:micrographs}b show optical images
of two $\alpha\textrm{-In}_{2}\textrm{Se}_{3}$ flakes. Each QL
is around 0.84\,nm thick, verified via atomic force microscopy. Due to the low
absorption of 1 QL and the weak optical contrast between the flakes and the
mica substrate, it was difficult to image the 1 QL regions with a normal
optical microscope. Since the visible light absorption increases linearly
with the thickness, larger QL stacks have better optical contrast, as can
be seen from the small center triangle with 5 QL in Figure
\ref{fig:micrographs}b.

Figures \ref{fig:micrographs}c and \ref{fig:micrographs}d show SHG micrographs
taken at a fundamental wavelength of 780\,nm (1.59\,eV), for the two samples
shown in Figures \ref{fig:micrographs}a and \ref{fig:micrographs}b,
respectively. The mica substrate is centrosymmetric, so we expect a strong SHG
response only from the $\alpha\textrm{-In}_{2}\textrm{Se}_{3}$ nano-flakes that
are non-centrosymmetric at room temperature. This provides better contrast
between the sample and the substrate. In Figures \ref{fig:micrographs}c and
\ref{fig:micrographs}d, we measured $s$-polarized SHG light generated by $s$-polarized incident electric fields.
We kept the laser intensity below
2\,mW to avoid laser-induced charge or damaging the sample. At this laser intensity,
the SHG response from 1 QL is very close to the background intensity. The SHG
response increases sharply from 1 QL to 2 QL; 3 QL has only a slightly higher
response than 2 QL. Note that the SHG intensity from 5 QL (small center triangle
in Figure \ref{fig:micrographs}d) is lower than that of 3 QL.
% {\color{blue}\textit{4 QL is also lower than 3 QL. It's just not shown in this micrograph.}}

For ease of notation, let us denote the incoming $1\omega$ photon polarization
as either $p$ or $s$, and analogously, the outgoing $2\omega$ photon
polarization with either capital $P$ or $S$. To elucidate the horizontal stacking
order of the two samples, we used the RASHG technique to measure SHG intensity 
with $sS$ polarization while rotating the sample. RASHG is
useful to confirm the relative crystallographic orientation between QLs, as well
as the symmetry relations (and thus space group) of the material. We used a
higher laser power to obtain a strong SHG response from 1 QL. We kept the laser
beam at one spot for no more than a second to avoid the spurious laser
induced effects. The $R3m$ space group has four independent nonzero components
of the nonlinear susceptibility tensor, $\chi^{xxx} = \chi^{xxy} = \chi^{yxy}$,
$\chi^{xxz} = \chi^{yyz}$, $\chi^{zxx} = \chi^{zyy}$, and $\chi^{zzz}$
\cite{boyd, popovbook}. From these components, we can fit the expected RASHG
patterns and determine the relative crystallographic orientation between
different QLs. We can isolate two tensor components, $\chi^{xxx}$ and
$\chi^{zxx}$, whose mirror plane is perpendicular to the $y$-axis, by selecting $sS$
and $sP$ polarization combinations. For these polarization configurations, the
outgoing SHG fields are related to the incoming fields as \cite{sipePRB87,
yamadaPRB94},
\begin{equation*}
\begin{split}
E_{S}(2\omega) &= 
    \left[
    A^{\prime}\sin3{\phi}\,\chi^{xxx}
    \right]E_{s}^{2}(\omega),\\
    % \label{eq:SHss}
E_{P}(2\omega) &=
    \left[
        A\chi^{zxx} +
        B\sin{3(\phi+\theta)}\chi^{xxx}
    \right]E_{s}^{2}(\omega),
    % \label{eq:SHsp}
\end{split}
\end{equation*}
where $A$, $B$, and $A^{\prime}$ are fitting parameters, $\phi$ is the azimuthal
angle and $\theta$ is the relative angle between QLs. $E_{s}(\omega)$ is the
electric field taken at a fundamental frequency $\omega$ that is incident on
the samples. We can readily see that $sS$ polarized SHG requires only $\chi^{xxx}$, while $sP$ polarized SHG depends on $\chi^{xxx}$ and $\chi^{zxx}$.
% (I moved this sentence here because the following paragraph is about the cyrs. orientation.)
% Later, we will extract $\chi^{xxx}$ and $\chi^{zxx}$ from these equations.
% (Okay, now we have to include the experimental $\chi^{xxx}$ and $\chi^{zxx}$}}

From these relations, we can determine the relative crystallographic orientation
between QLs. For the sample shown on the left side of Figure
\ref{fig:micrographs} (a, c, and e) , all QLs are oriented along the same
direction. On the other hand, the sample shown on the right side of Figure
\ref{fig:micrographs} (b, d, and f), has its 1 QL region rotated by 34$^{\circ}$
with respect to the overlaying QLs. As seen in Figure \ref{fig:micrographs}f,
the maxima of the 1 QL RASHG values (black crosses) fall between the maxima of
the 2 QL (red exes) and 3 QL (blue circles). From this point on, we will refer
the left sample as $\theta = 0^{\circ}$ (also known as AB stacking order), the
right sample as $\theta = 34^{\circ}$. We verified the crystallographic
orientation of more than 15 other nano-flakes on the same substrate, with all of
them having either a value of $\theta = 0^{\circ}$ or $\theta = 180^{\circ}$;
these samples are even discernible via conventional optical microscopy.

Although $\theta = 34^{\circ}$ is not a preferred stacking orientation, it is
possible that a defect on the first QL acted as a seed for the next layers during
the growth process, which made it grow at a random crystallographic orientation.
It is clear from Figures \ref{fig:micrographs}c and \ref{fig:micrographs}d that
this stacking angle strongly affects the SHG responses; the overall SHG
intensity of $\theta = 34^{\circ}$ at 780\,nm is weaker by a factor of three
compared to $\theta = 0^{\circ}$ for both $sS$ and $sP$ SHG polarizations.
Similar trends have been observed on artificially stacked MoS$_{2}$ layers; the
Raman, SHG, and photoluminescence spectra were greatly affected by the
horizontal stacking orientation due to changes in the interlayer coupling
\cite{plechinger2DM15}. Therefore, we can expect that the interlayer distance
between 1 and 2 QL is larger when $\theta = 34^{\circ}$ than the interlayer distance for
$\theta = 0^{\circ}$, which leads to a weaker interlayer coupling.

\begin{figure}[t]
\centering
\includegraphics[width=\linewidth]{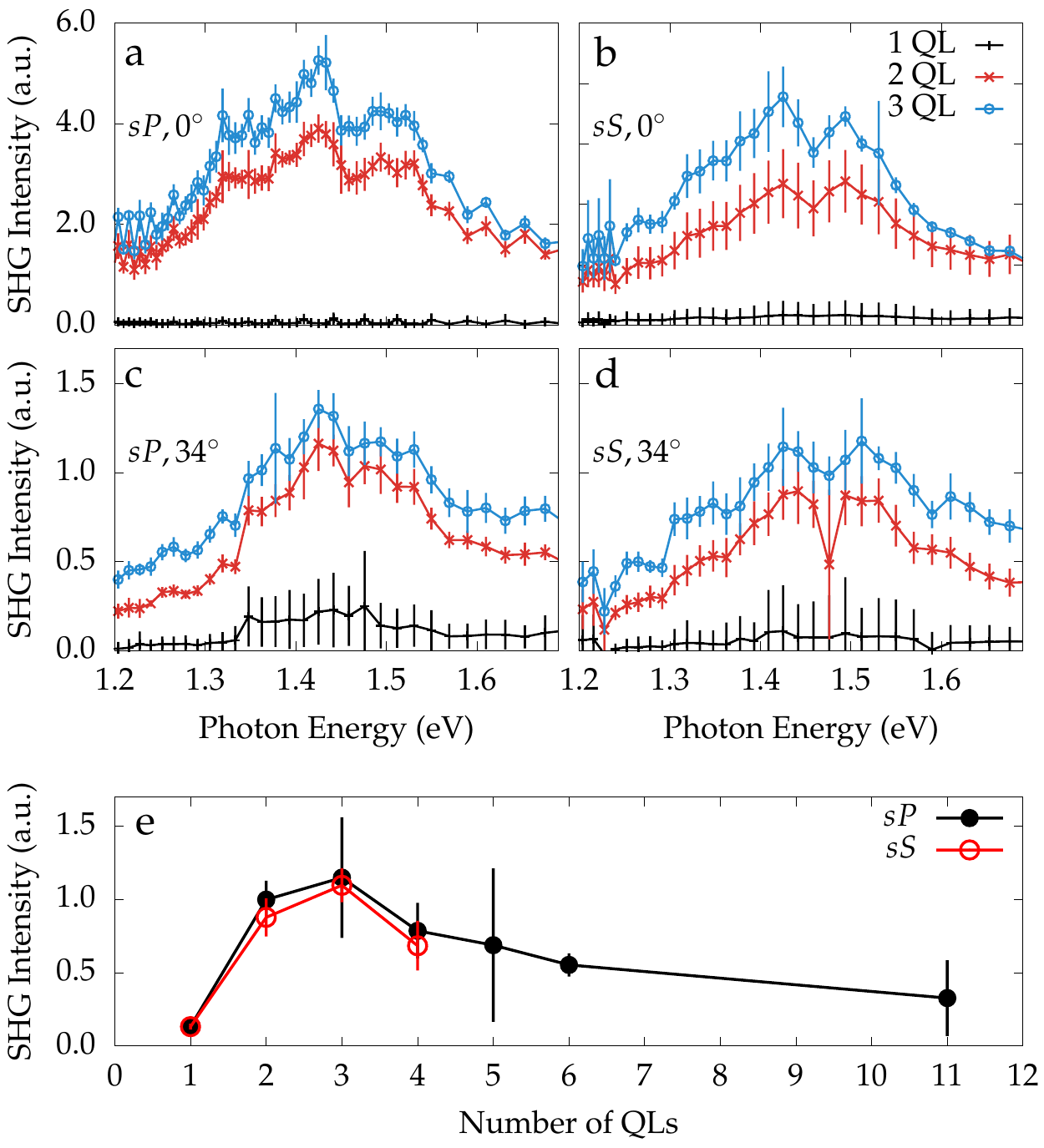}
\caption{Normalized SHG intensity for: \textbf{a}, $sP$ polarization for the
$\theta = 0^{\circ}$ sample; \textbf{b}, $sS$ polarization for the $\theta =
0^{\circ}$ sample; \textbf{c} $sP$ polarization for the $\theta = 34^{\circ}$
sample; and, \textbf{c} $sS$ polarization for the $\theta = 34^{\circ}$.
\textbf{e}, normalized SHG intensity as a function of the number of layers,
measured at 780\,nm (1.59\,eV) fundamental wavelength for $sP$ and $sS$
polarizations for the $\theta = 0^{\circ}$ sample. All SHG intensities were
normalized against $\alpha$-quartz.}
\label{fig:spectra}
\end{figure}

To observe the spectroscopic SHG response, we varied the fundamental wavelength
from 1.2 to 1.7\,eV (730 to 1040\,nm). We fixed the azimuthal angle to a maximum
in the RASHG measurement for each polarization configuration of Figure
\ref{fig:micrographs}.
% which reduces equations \ref{eq:SHss} and \ref{eq:SHsp} to
% \begin{align}
% E_{S}(2\omega) &= A^{\prime}\chi^{xxx}E_{s}^{2}(\omega),\\
% E_{P}(2\omega) &= \left(A\chi^{zxx} + B\chi^{xxx}\right)E_{s}^{2}(\omega). 
% \end{align}
Figures \ref{fig:spectra}a--\ref{fig:spectra}d presents the SHG spectra measured
for $sP$ and $sS$ polarizations for both $\theta = 0^{\circ}$ and $\theta =
34^{\circ}$ samples. Both samples present a broad resonance around 1.4\,eV
originating from a resonant optical transition around the $\Gamma$ point, with
$sP$ polarization yielding a more intense SHG response than $sS$. For $\theta = 0^{\circ}$, the $sP$ SHG intensity around 1.4\,eV is around twice as large as
the $sS$ intensity, and $\theta = 0^{\circ}$ is 3 times more intense than $\theta =
34^{\circ}$ for $sP$ polarization, and around 2 times more for $sS$
polarization. Although 1 QL appears very small, we know from Figure
\ref{fig:micrographs} that 2 QL response is around 5 times more intense. 3 QL
intensity is around 25\,\% (18\,\%) more intense than 2 QL for $\theta =
0^{\circ}$ ($\theta = 34^{\circ}$).
% We also observed that 4 QL has lower SHG response than 3 QLs.
Figure \ref{fig:spectra}e presents the normalized SHG intensity as a function
the number of QLs at 1.59\,eV. We normalized the intensity with respect to the 2 QL response
. Above 3 QLs, the SHG intensity decreases as the number of layers
increases. This experimentally validates that the net polarization (electric
dipole) does not increase monotonically with the number of QLs; instead, some
amount of out-of-plane polarization is canceled out due to charge transfer, generating
opposite polarization across the film \cite{dingNC17}. From those two samples, it is clear that the
SHG response of $\alpha\textrm{-In}_{2}\textrm{Se}_{3}$ has not only thickness
dependence but is also highly sensitive to the horizontal stacking arrangement.

% \begin{figure*}[t]
% \centering
% \includegraphics[width=0.9\linewidth]{fig03}
% \caption{}
% \label{fig:spectra1} 
% \end{figure*}

As shown previously, s-in/s-out SHG intensity depends on one nonlinear
tensor component, $\chi_{xxx}^{(2\omega)}$, and s-in/p-out depends on $\chi_{xxx}^{(2\omega)}$ and 
$\chi_{zxx}^{(2\omega)}$. Thus, we calculated $\chi_{xxx}^{(2\omega)}$ from s-in/s-out SHG 
intensity and estimated $\chi_{zxx}^{(2\omega)}$ using $\chi_{xxx}^{(2\omega)}$. The equations
can be rewritten as:
\begin{align}
\left\vert\chi^{xxx}\right\vert &= 
    \sqrt{I_{sS}}\,\frac{1}{A^{\prime}}\frac{E^{Q}}{E_{s}^{2}},\\
\left\vert\chi^{zxx}\right\vert &=
    \sqrt{I_{sP}}\,\frac{1}{A}\frac{E^{Q}}{E_{s}^2} - \frac{B}{A}
    \left\vert\chi^{xxx}\right\vert
\end{align}
where $E^Q$ indicates the SH fields generated from $\alpha$-quartz, a reference sample in the
experiment. $A^{\prime}$, $A$, and $B$ are the geometric factors, including refractive indices at 
$\omega$, $2\omega$ of In$_2$Se$_3$, and the angle of incidence. We calculated them to extract 
the magnitudes of the $\chi_{xxx}^{(2\omega)}$ and $\chi_{zxx}^{(2\omega)}$ (see Supporting 
Material for details). We used the refactive index from the bulk $\alpha\textrm{-In}_{2}\textrm{Se}_{3}$
to have a wide spectral range as well as to stay less sensitive to the ellipsometry fitting model.
Our DFT calculation expects the refractive indices from the bulk and a few layers to be off by less than 15\,\% in magnitude with similar spectral shape.
% {\color{red}can we add something 
% to support this argument?

The relative magnitudes of the tensor components are consistent between the experiment and the 
theory. In particular, sharp increase between 1QL and 2 QL has been observed which might be
related to the confinement effect in the 2D limit and the electric dipole arranement. The tensor 
components on 3 QL is still larger than the others throughout the measured photon energy range 
by less than 20\%. $\chi_{xxx}$ is smaller than $\chi_{zxx}$ in magnitude by a factor of 3 in 2 QL 
and 4 in 3 QL. 

Since the phase difference between $\chi_{xxx}^{(2\omega)}$ and $\chi_{zxx}^{(2\omega)}$
could not be determined directly in the experiment, when we calculated $\chi_{zxx}$ from 
s-in/p-out SHG intensity, we calculated both \textit{in-phase} and \textit{out-of-phase} cases. 
When the two components are \textit{in-phase}, the relative magnitudes
differ by a couple of orders of magnitude, while they are in the same orders of magnitudes
when \textit{out-of-phase}. The latter case agrees with our DFT calculation and therefore, 
we concluded that the two components are close to \textit{out-of-phase}.

\subsection{Transmission and Dielectric Functions}

%{\color{blue}\textit{How about having a sentence before this section
%to connect it to the part B?}}

\begin{figure}[t]
\centering
\includegraphics[width=\linewidth]{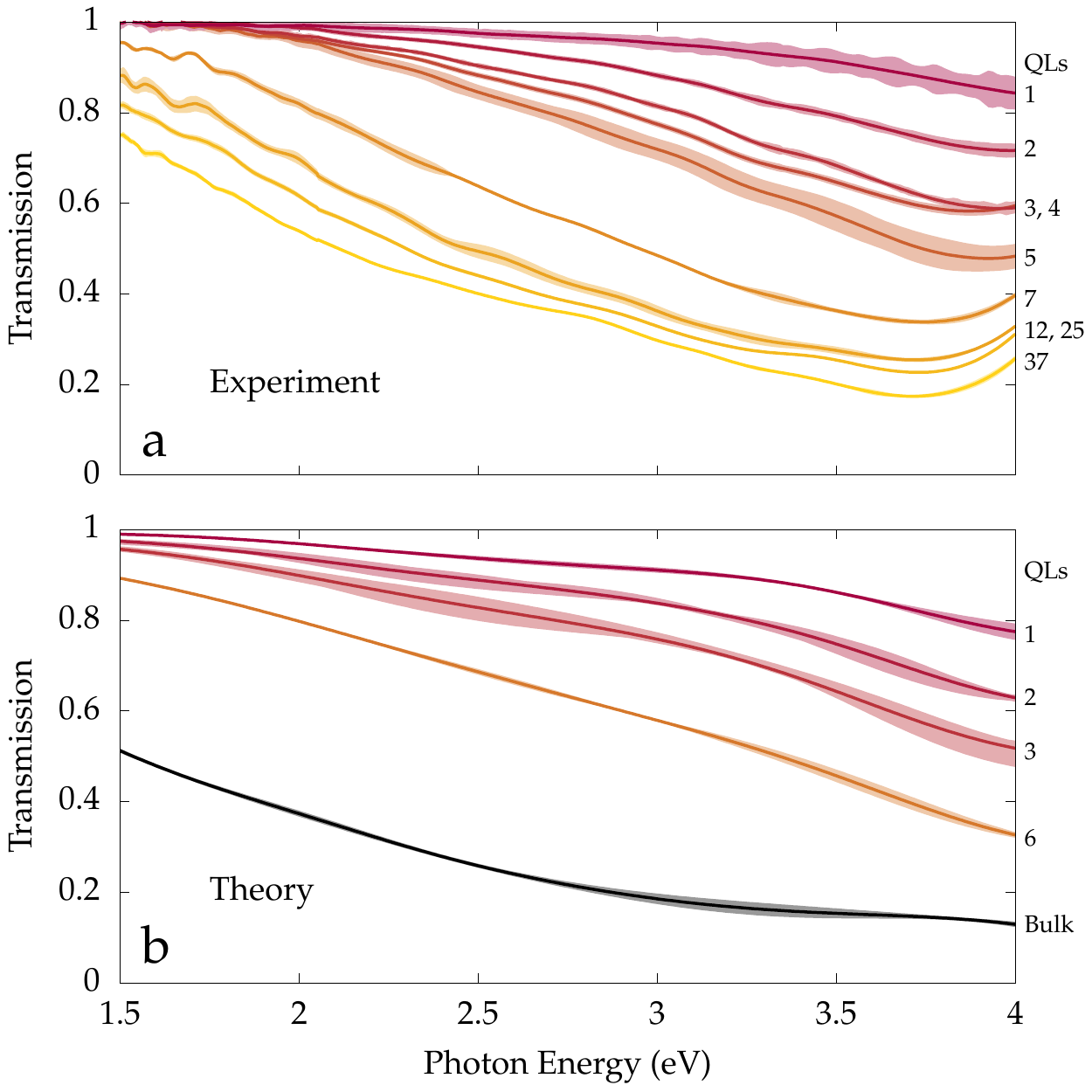}
\caption{\textbf{a}, experimental transmission spectra for layered
$\alpha\textrm{-In}_{2}\textrm{Se}_{3}$ in the range of 1.5--4\,eV, for 1 to 37
QLs. \textbf{b}, theoretical results for 1--3 QLs, 6 QLs and bulk. Shaded
regions around the theoretical curves represent the different stacking
arrangements for any given number of QLs.}
\label{fig:trans} 
\end{figure}

Fig. \ref{fig:trans}a presents the measured transmission spectra of our
samples for 1--5, 7, 12, 25, and 37 QLs, over an energy range of 1.5 to 4\,eV.
We were unable to find a 6 QL region in the synthesized samples. At each photon
energy, total sample transmission was divided by the separately measured
transmission of the bare mica substrate, in order to isolate the optical
response of the nanofilm. The shaded region represents the measurement
uncertainty. As mentioned previously, we calculated the optical response for two
representative structures (wurtzite-like and zincblende-like), along with many
possible stacking arrangements that are possible for two or more QLs, including
different horizontal placement between QLs (with no relative rotation), and various combinations of dipole
directions for individual QLs in a stack. Our calculated transmission spectra
are presented in Figure \ref{fig:trans}b for 1--3 QLs, 6 QLs, and bulk. The
shaded regions around each curve correspond to the maximum and minimum values of
the numerous stacking arrangements, while the solid curves represent the average
value for each number of QLs. Calculations of the optical response for more than
6 QLs, or $G_{0}W_{0}$ self-corrections on the bandgap for more than 2 QLs,
were not practical due to the required computational resources. Therefore, these
theoretical results have no quasiparticle correction applied to them. Calculated
band structures (not shown) indicate that all stacking arrangements have
indirect band gaps.

We first note that the various stacking arrangements yield calculated spectra
that are quite consistent for each number of QLs. For 1-3 QLs, calculated and
measured transmission differ by around 6\,\% over the measured spectral range.
The calculated spectrum for 6 QLs also differ by less than 7\,\% from the
average transmission of the 5 and 7 QL samples. For these cases, the
calculations reproduced the observed decrease in transmission from 1.5 to 4 eV
reasonably well. The experimental transmission spectra below 7 QLs are dominated
by a single broad peak around 4\,eV and a small shoulder around 2.5\,eV. Above 7
QLs, the lower energy peak becomes more prominent, and shifts to $\sim 2.3$\,eV,
while the higher energy peak shifts to $\sim$ 3.7 eV. For the calculated spectra,
the lower energy peak manifests itself as a tail or shoulder in the calculated
spectra. In bulk $\alpha\textrm{-In}_{2}\textrm{Se}_{3}$, it becomes a more
prominent shoulder. The peak above 3.4\,eV (2.5\,eV) corresponds to optical
transitions around the K (M) points. Thus hereafter, we will call it the K- (M-)
peak. We have not applied any form of quasiparticle energy
correction to the theoretical results for these layered samples. This tends to
increase the calculated bandgap, which blue-shifts the dielectric function, and
consequently, the calculated transmission spectra. However, even without the
corrected eigen-energies, our calculations not only describe the trends in the
experimental data, but are also consistent with previously reported experimental
\cite{queredaAOM16} and theoretical \cite{huRA17} results.

\begin{figure}[t]
\centering
\includegraphics[width=\linewidth]{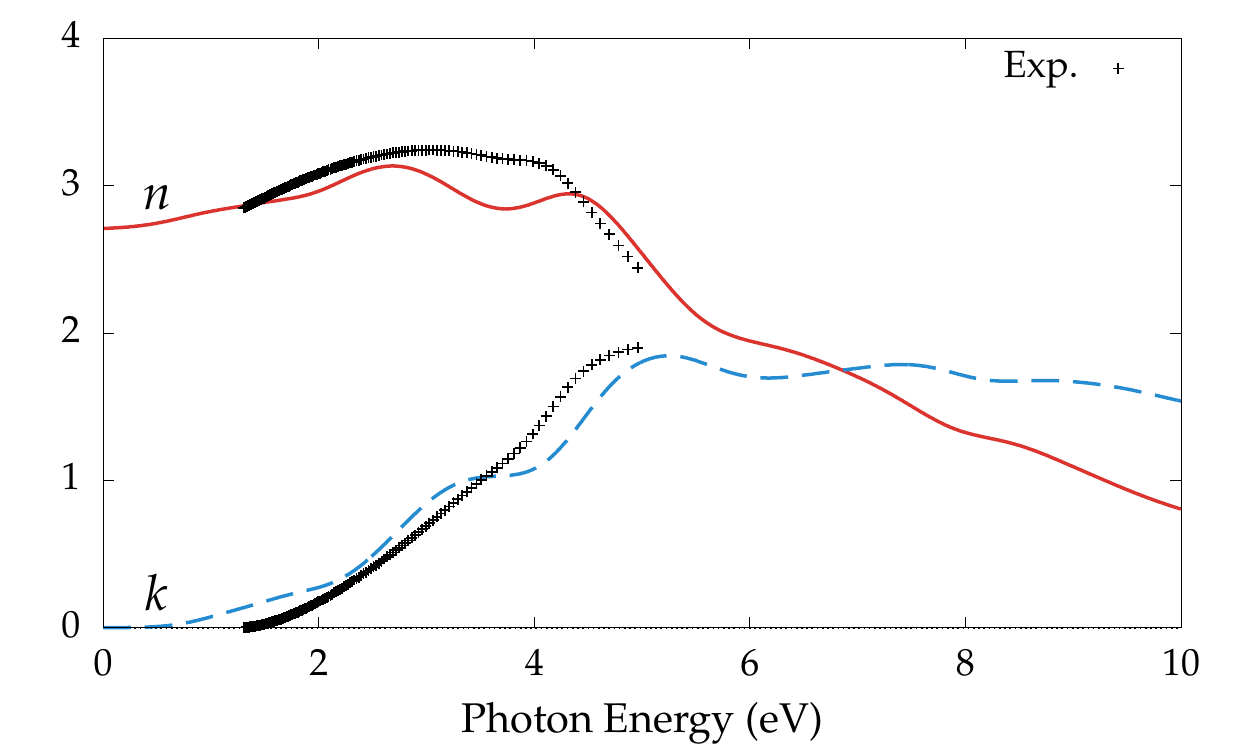}
\caption{The real and imaginary parts of the
complex index of refraction ($n$ and $k$) for bulk
$\alpha\textrm{-In}_{2}\textrm{Se}_{3}$, calculated over a photon energy range
of 0--10\,eV and measured in 1.5--5\,eV. Theoretical curves use $G_{0}W_{0}$
obtained quasiparticle correction for the band gap of 1.11\,eV.}
\label{fig:eblk}
\end{figure}

Fig. \ref{fig:eblk} compares the experimental (1.5--5\,eV) and theoretical
(0--10\,eV) refractive indices for bulk $\alpha\textrm{-In}_{2}\textrm{Se}_{3}$.
The scatter points are from the experimental SE fitting and the solid curves are
from TDDFT calculations. We used the high spatial resolution ($\sim
1\,\mu\mathrm{m}$) of the imaging ellipsometer to avoid taking data from the
sample edges and the adsorbed dirt from the air. To obtain the ellipsometric
dielectric functions, we constructed a model using two Tauc-Lorentz oscillators.
Mean-Square-Error in the fitting was less than 15. The theoretical curves in
figure \ref{fig:eblk} include a quasiparticle correction of 0.66\,eV, obtained
from an \emph{ab initio} $G_{0}W_{0}$ calculation, placing the optical band gaps
at 1.11\,eV. This value differs by 5\,\% from the experimental value
of 1.167\,eV, and are slightly lower than previously reported values for the
$G_{0}W_{0}$ corrected optical band gap \cite{queredaAOM16}. The calculated $k$
spectrum has a dominant peak around 4.7\,eV, with a smaller shoulder at around
3.35\,eV. Although these peaks are not resolved experimentally, the measured
real $n$ (imaginary $k$) differ on average by less than 15\,\% (30\,\%)
throughout the measured photon energy range.

The calculated band structure (not shown) is consistent with previously
reported results \cite{queredaAOM16}, with the optical band gaps occurring
around the $\Gamma$ point. Transitions begin to accumulate rapidly around
2.5\,eV (isolated around the M point). The greatest density of transitions
occurs between 3.0 and 4.0 eV, primarily between the $\Gamma$, K, and M points.
Transitions above 4.0\,eV correspond to transitions between the second-highest
valence band and the lowest conduction band, or between the highest valence band
and the third lowest conduction band.

Although we cannot discern between the theoretical structures or between the
numerous possible stacking configurations from this comparison of linear optical
quantities, the overall agreement between theory and experiment is a strong
indicator that the selected theoretical structures represent the atomic
configuration of the real samples. The results presented above provide two
important points of comparison. First, excellent agreement for the bulk optical
constants, and good agreement for the dielectric functions for the layered
samples both demonstrate that the intrinsic material properties are accurately
portrayed by our combination of relaxed atomic coordinates and \emph{ab initio}
calculations. Second, agreement between the experimental transmission curves
(via direct measurement) compared with the calculated dielectric functions
(within a straightforward transmission model) indicates that the intrinsic
material properties are accurately matched with real experimental conditions,
such as individual QL thickness.

%%%%%%%%%%%%%%%%%%%%%%%%%%%%%%%%%%%%%%%%%%%%%%%%%%%%%%%%%%%%%%%%%%%%%%%%%%%%%%%%
%%%%%%%%%%%%%%%%%%%%%%%%%%%%%%%%%%%%%%%%%%%%%%%%%%%%%%%%%%%%%%%%%%%%%%%%%%%%%%%%

\subsection{Band Gaps}

\begin{figure}[b]
\centering
\includegraphics[width=\linewidth]{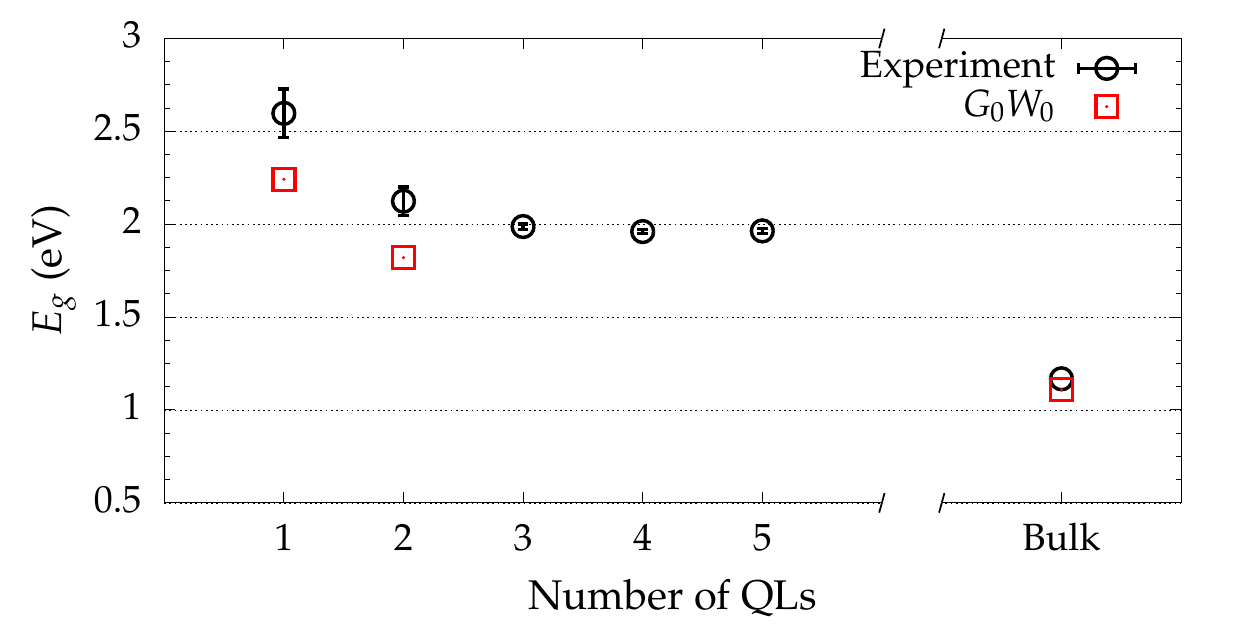}
\caption{The optical band gap values obtained from the experimental measurements, and
the two corresponding bulk values obtained from a $G_{0}W_{0}$ calculation.}
\label{fig:gaps} 
\end{figure}

Fig. \ref{fig:gaps} presents the optical band gap dependence on the number of
QLs, extracted from the transmission curves for the layered samples, and ellipsometry fitting for the bulk.
%{\color{blue}\textit{We calculated the corresponding absorption spectra using 
%Beer-Lambert law.}}
To obtain the error bars, we
repeated the fitting on several measurements obtained from different spots on
the same flake. The values decrease steadily as the number of layers increases,
starting at $\sim$2.5\,eV for 1 QL down to $\sim$1.167\,eV for the bulk. The
latter value is lower than the previously reported values ($\sim$1.4\,eV)
obtained from a transmission measurement and \emph{ab initio} $G_{0}W_{0}$
calculations \cite{debbichiJPCL15, queredaAOM16}, although it is well within the
range of the reported optical band gaps for
$\alpha\textrm{-In}_{2}\textrm{Se}_{3}$ (see Table \ref{tab:bandgaps}). Our
$G_{0}W_{0}$ calculation yields a bulk optical band gap value of 1.11\,eV, which
agrees within 5\,\% of the measured optical band gap; for 1 and 2 QLs, the
calculated bandgap differs by 15\,\% of the measured values. Table
\ref{tab:bandgaps} present a comprehensive summary of reported optical bandgap
values, including those presented in this work. As we can see from the table,
the gap values (both experimental and theoretical) span a wide range; in bulk,
the optical band gap varies from 1.154\,eV to 1.48\,eV. However, the band gaps
reported in this work are consistent with the ones available from the literature.

\begin{table}
\caption{The optical band gaps (in eV) presented in this work compared to values
reported in the literature.}
\label{tab:bandgaps}
\begin{ruledtabular}
\begin{tabular}{r c c }
Sample  & Experiment (eV) & $G_{0}W_{0}$ (eV)\\
\hline
1 QL    & $2.597\pm0.130$ & 2.241 \\
2 QL    & $2.123\pm0.075$ & 1.820 \\
3 QL    & $1.987\pm0.014$ & --    \\
4 QL    & $1.960\pm0.010$ & --    \\
5 QL    & $1.963\pm0.013$ & --    \\
Bulk    &  1.167          & 1.110 \\
\end{tabular}
\end{ruledtabular}
\end{table}

The motion of electrons in one or multiple QLs is confined to the direction
perpendicular to the QL plane, while the carriers can move freely along the
films. Such confinement of the charge carriers can be qualitatively explained
considering that the $\textrm{In}_{2}\textrm{Se}_{3}$ flakes act as 1D quantum
wells \cite{queredaAOM16}. In terms of the electron band structure, if the
corresponding 3D layered system is semiconducting with banggap $E_{g}$, the
corresponding free-standing or deposited 2D layers (especially on an insulating
substrate) should have a larger bandgap due to this confinement effect. The
largest increase in the band gap occurs when going from one to multiple QLs, as
the value decreases inversely proportional to the square of the film thickness
\cite{queredaAOM16}. This simple model, which also depends on the effective mass
of the electrons and holes, yield an energy decrease of 0.47\,eV when going from
one QL to two QLs, and a decrease of 0.14\,eV between two QLs and three QLs. The
gap widening in nano-thin films with decreasing thickness is clearly observed in
our first principle calculations \cite{shkrebtiiproc}.
% It is worth noting that
% the actual modification of the gap due to confinement, calculated from first
% principles, does not exactly follow the qualitative 1D quantum well model.
% {\color{blue}\textit{Earlier in this paragraph said In2Se3 can act as 1D QWs but
% why doesn't it follow the qualitative 1D QW model?}}

%%%%%%%%%%%%%%%%%%%%%%%%%%%%%%%%%%%%%%%%%%%%%%%%%%%%%%%%%%%%%%%%%%%%%%%%%%%%%%%%
%%%%%%%%%%%%%%%%%%%%%%%%%%%%%%%%%%%%%%%%%%%%%%%%%%%%%%%%%%%%%%%%%%%%%%%%%%%%%%%%

\section{Conclusion}\label{sec:conclusion}

We have investigated the linear and nonlinear optical response of a few layers of
$\alpha\textrm{-In}_{2}\textrm{Se}_{3}$. Spectroscopic transmission and ellipsometry
measurements were used to obtain the linear optical response. These two
independent methods, combined with \emph{ab initio} calculations for 1-3, 6 QL and
bulk, provide a comprehensive set of optical response from single QL to bulk
$\alpha\textrm{-In}_{2}\textrm{Se}_{3}$. Our calculations provide good agreement
with the experimental data; in particular $G_{0}W_{0}$ calculation yields a bandgap 
for the bulk that is within 10\'\% of the experimental values. Calculated
TDDFT and DFT-LDA \emph{ab initio} dielectric functions and transmission spectra
are comparable to those obtained from spectroscopic ellipsometry and
transmission measurements.

%We have investigated the linear optical response of layered
%$\alpha\textrm{-In}_{2}\textrm{Se}_{3}$ from one quintuple layer (QL) up to
%bulk, using spectroscopic transmission and ellipsometry measurements. These two
%independent methods, combined with \emph{ab initio} calculations for 1-6 QL and
%bulk, provide a comprehensive set of optical constants from single QL to bulk
%$\alpha\textrm{-In}_{2}\textrm{Se}_{3}$. Our calculations provide good agreement
%with the experimental data; in particular $G_{0}W_{0}$ calculation yields a bandgap 
%for the bulk that is within 10\'\% of the experimental values. Calculated
%TDDFT and DFT-LDA \emph{ab initio} dielectric functions and transmission spectra
%are comparable to those obtained from spectroscopic ellipsometry and
%transmission measurements.

From the experimental data, we analyzed two peaks around 2.5\,eV and 4\,eV,
which correspond to optical transitions around the M- and K-points in momentum
space, respectively. We have shown that changing the thickness of the material
tunes the optical band gap of $\alpha\textrm{-In}_{2}\textrm{Se}_{3}$ from
visible (2.6\,eV) to infrared (1.17\,eV) photon energies, with significant
absorption. The optical band gap tunability range of 1.4\,eV is consistent with
previous measurements for QL thicknesses between 3--25\,nm \cite{queredaAOM16}
and is one of the largest amongst known 2D materials, which typically ranges
$\sim$0.5\,eV \cite{queredaAOM16, gusakovaPSSA17}. This property suggests
potential applications as photodetectors \cite{zhengNC15, zhengJMCC16} or as
solar cell material. 

Although linear optical measurements show how the
dielectric functions change with photon energy and QL thickness, it is insensitive to
the change in stacking order or net polarization, which is critical to understand ferroelectricity
on a few QLs of $\alpha\textrm{-In}_{2}\textrm{Se}_{3}$. 
We used second-harmonic generation microscopy and spectroscopy to probe it.
The SHG response peaks at 3 QL and decreases at thicker QLs, as oppose to 
monotonically changing linear response, indicating that 3 QL has the strongest
net polarization. In addition, the horizontal stacking order significantly changes
the SHG response; by rotating $\sim30^{\circ}$, the SHG response is decreased
by a factor of 3. This tunability suggests that we can fine-tune the optical response
towards desired optoelectronic applications.

%From the experimental data, we analyzed two peaks around 2.5\,eV and 4\,eV,
%which correspond to optical transitions around the M- and K-points in momentum
%space, respectively. We have shown that changing the thickness of the material
%tunes the optical band gap of $\alpha\textrm{-In}_{2}\textrm{Se}_{3}$ from
%visible (2.54\,eV) to infrared (1.17\,eV) photon energies, with significant
%absorption. The optical band gap tunability range of 1.4\,eV is consistent with
%previous measurements for QL thicknesses between 3--25\,nm \cite{queredaAOM16}
%and is one of the largest amongst known 2D materials, which typically ranges
%$\sim$0.5\,eV \cite{queredaAOM16, gusakovaPSSA17}. This property suggests
%potential applications as photodetectors \cite{zhengNC15, zhengJMCC16} or as
%solar cell material. Although linear optical measurements show how the
%dielectric functions change with photon energy and QL thickness, to observe
%changes due to net polarization or stacking order, nonlinear optical methods
%such as second-harmonic generation microscopy and spectroscopy are better suited
%because of their sensitivity to the net polarization.

%%%%%%%%%%%%%%%%%%%%%%%%%%%%%%%%%%%%%%%%%%%%%%%%%%%%%%%%%%%%%%%%%%%%%%%%%%%%%%%%

\section{Acknowledgments}

This work was supported by Robert Welch Foundation grant F-1038. Y. Cho
acknowledges InProTUC (International Promovieren an der Technischen Universität
Chemnitz) program for supporting her research at TU Chemnitz. Sample preparation
was supported by Welch Foundation Grant F-1814, and the U.S. Department of
Energy (DOE), Office of Science, Basic Energy Sciences, under Awards
DE-SC0010308 and DE-SC0019025. Additionally, the authors thankfully acknowledge
the computer resources, technical expertise, and support provided by the
Laboratorio Nacional de Superc\'omputo del Sureste de M\'exico, a member of the
CONACYT network of national laboratories as well as the Shared Hierarchical
Academic Research Computing Network (SHARCNET) of Ontario, Canada.

%%%%%%%%%%%%%%%%%%%%%%%%%%%%%%%%%%%%%%%%%%%%%%%%%%%%%%%%%%%%%%%%%%%%%%%%%%%%%%%%
%%%%%%%%%%%%%%%%%%%%%%%%%%%%%%%%%%%%%%%%%%%%%%%%%%%%%%%%%%%%%%%%%%%%%%%%%%%%%%%%

\bibliographystyle{apsrev4-1}

\begin{thebibliography}{57}%
\makeatletter
\providecommand \@ifxundefined [1]{%
 \@ifx{#1\undefined}
}%
\providecommand \@ifnum [1]{%
 \ifnum #1\expandafter \@firstoftwo
 \else \expandafter \@secondoftwo
 \fi
}%
\providecommand \@ifx [1]{%
 \ifx #1\expandafter \@firstoftwo
 \else \expandafter \@secondoftwo
 \fi
}%
\providecommand \natexlab [1]{#1}%
\providecommand \enquote  [1]{``#1''}%
\providecommand \bibnamefont  [1]{#1}%
\providecommand \bibfnamefont [1]{#1}%
\providecommand \citenamefont [1]{#1}%
\providecommand \href@noop [0]{\@secondoftwo}%
\providecommand \href [0]{\begingroup \@sanitize@url \@href}%
\providecommand \@href[1]{\@@startlink{#1}\@@href}%
\providecommand \@@href[1]{\endgroup#1\@@endlink}%
\providecommand \@sanitize@url [0]{\catcode `\\12\catcode `\$12\catcode
  `\&12\catcode `\#12\catcode `\^12\catcode `\_12\catcode `\%12\relax}%
\providecommand \@@startlink[1]{}%
\providecommand \@@endlink[0]{}%
\providecommand \url  [0]{\begingroup\@sanitize@url \@url }%
\providecommand \@url [1]{\endgroup\@href {#1}{\urlprefix }}%
\providecommand \urlprefix  [0]{URL }%
\providecommand \Eprint [0]{\href }%
\providecommand \doibase [0]{http://dx.doi.org/}%
\providecommand \selectlanguage [0]{\@gobble}%
\providecommand \bibinfo  [0]{\@secondoftwo}%
\providecommand \bibfield  [0]{\@secondoftwo}%
\providecommand \translation [1]{[#1]}%
\providecommand \BibitemOpen [0]{}%
\providecommand \bibitemStop [0]{}%
\providecommand \bibitemNoStop [0]{.\EOS\space}%
\providecommand \EOS [0]{\spacefactor3000\relax}%
\providecommand \BibitemShut  [1]{\csname bibitem#1\endcsname}%
\let\auto@bib@innerbib\@empty
%</preamble>
\bibitem [{\citenamefont {Boukhvalov}\ \emph {et~al.}(2017)\citenamefont
  {Boukhvalov}, \citenamefont {Gürbulak}, \citenamefont {Duman}, \citenamefont
  {Wang}, \citenamefont {Politano}, \citenamefont {Caputi}, \citenamefont
  {Chiarello},\ and\ \citenamefont {Cupolillo}}]{boukhvalovNM17}%
  \BibitemOpen
  \bibfield  {author} {\bibinfo {author} {\bibfnamefont {D.}~\bibnamefont
  {Boukhvalov}}, \bibinfo {author} {\bibfnamefont {B.}~\bibnamefont
  {Gürbulak}}, \bibinfo {author} {\bibfnamefont {S.}~\bibnamefont {Duman}},
  \bibinfo {author} {\bibfnamefont {L.}~\bibnamefont {Wang}}, \bibinfo {author}
  {\bibfnamefont {A.}~\bibnamefont {Politano}}, \bibinfo {author}
  {\bibfnamefont {L.}~\bibnamefont {Caputi}}, \bibinfo {author} {\bibfnamefont
  {G.}~\bibnamefont {Chiarello}}, \ and\ \bibinfo {author} {\bibfnamefont
  {A.}~\bibnamefont {Cupolillo}},\ }\href {\doibase 10.3390/nano7110372}
  {\bibfield  {journal} {\bibinfo  {journal} {Nanomaterials}\ }\textbf
  {\bibinfo {volume} {7}},\ \bibinfo {pages} {372} (\bibinfo {year}
  {2017})}\BibitemShut {NoStop}%
\bibitem [{\citenamefont {Demirci}\ \emph {et~al.}(2017)\citenamefont
  {Demirci}, \citenamefont {Avazl{\i}}, \citenamefont {Durgun},\ and\
  \citenamefont {Cahangirov}}]{demirciPRB17}%
  \BibitemOpen
  \bibfield  {author} {\bibinfo {author} {\bibfnamefont {S.}~\bibnamefont
  {Demirci}}, \bibinfo {author} {\bibfnamefont {N.}~\bibnamefont {Avazl{\i}}},
  \bibinfo {author} {\bibfnamefont {E.}~\bibnamefont {Durgun}}, \ and\ \bibinfo
  {author} {\bibfnamefont {S.}~\bibnamefont {Cahangirov}},\ }\href {\doibase
  10.1103/PhysRevB.95.115409} {\bibfield  {journal} {\bibinfo  {journal} {Phys.
  Rev. B}\ }\textbf {\bibinfo {volume} {95}},\ \bibinfo {pages} {115409}
  (\bibinfo {year} {2017})}\BibitemShut {NoStop}%
\bibitem [{\citenamefont {Lin}\ \emph {et~al.}(2018)\citenamefont {Lin},
  \citenamefont {Fang}, \citenamefont {Tao}, \citenamefont {Li}, \citenamefont
  {Huang}, \citenamefont {Ding}, \citenamefont {Huang},\ and\ \citenamefont
  {Zhang}}]{linCEC18}%
  \BibitemOpen
  \bibfield  {author} {\bibinfo {author} {\bibfnamefont {J.}~\bibnamefont
  {Lin}}, \bibinfo {author} {\bibfnamefont {Z.}~\bibnamefont {Fang}}, \bibinfo
  {author} {\bibfnamefont {H.}~\bibnamefont {Tao}}, \bibinfo {author}
  {\bibfnamefont {Y.}~\bibnamefont {Li}}, \bibinfo {author} {\bibfnamefont
  {X.}~\bibnamefont {Huang}}, \bibinfo {author} {\bibfnamefont
  {K.}~\bibnamefont {Ding}}, \bibinfo {author} {\bibfnamefont {S.}~\bibnamefont
  {Huang}}, \ and\ \bibinfo {author} {\bibfnamefont {Y.}~\bibnamefont
  {Zhang}},\ }\href {\doibase 10.1039/C8CE00154E} {\bibfield  {journal}
  {\bibinfo  {journal} {CrystEngComm}\ }\textbf {\bibinfo {volume} {20}},\
  \bibinfo {pages} {2573} (\bibinfo {year} {2018})}\BibitemShut {NoStop}%
\bibitem [{\citenamefont {Henck}\ \emph {et~al.}(2019)\citenamefont {Henck},
  \citenamefont {Pierucci}, \citenamefont {Zribi}, \citenamefont {Bisti},
  \citenamefont {Papalazarou}, \citenamefont {Girard}, \citenamefont {Chaste},
  \citenamefont {Bertran}, \citenamefont {Le~F{\`e}vre}, \citenamefont
  {Sirotti}, \citenamefont {Perfetti}, \citenamefont {Giorgetti}, \citenamefont
  {Shukla}, \citenamefont {Rault},\ and\ \citenamefont {Ouerghi}}]{henckPRM19}%
  \BibitemOpen
  \bibfield  {author} {\bibinfo {author} {\bibfnamefont {H.}~\bibnamefont
  {Henck}}, \bibinfo {author} {\bibfnamefont {D.}~\bibnamefont {Pierucci}},
  \bibinfo {author} {\bibfnamefont {J.}~\bibnamefont {Zribi}}, \bibinfo
  {author} {\bibfnamefont {F.}~\bibnamefont {Bisti}}, \bibinfo {author}
  {\bibfnamefont {E.}~\bibnamefont {Papalazarou}}, \bibinfo {author}
  {\bibfnamefont {J.-C.}\ \bibnamefont {Girard}}, \bibinfo {author}
  {\bibfnamefont {J.}~\bibnamefont {Chaste}}, \bibinfo {author} {\bibfnamefont
  {F.}~\bibnamefont {Bertran}}, \bibinfo {author} {\bibfnamefont
  {P.}~\bibnamefont {Le~F{\`e}vre}}, \bibinfo {author} {\bibfnamefont
  {F.}~\bibnamefont {Sirotti}}, \bibinfo {author} {\bibfnamefont
  {L.}~\bibnamefont {Perfetti}}, \bibinfo {author} {\bibfnamefont
  {C.}~\bibnamefont {Giorgetti}}, \bibinfo {author} {\bibfnamefont
  {A.}~\bibnamefont {Shukla}}, \bibinfo {author} {\bibfnamefont {J.~E.}\
  \bibnamefont {Rault}}, \ and\ \bibinfo {author} {\bibfnamefont
  {A.}~\bibnamefont {Ouerghi}},\ }\href {\doibase
  10.1103/PhysRevMaterials.3.034004} {\bibfield  {journal} {\bibinfo  {journal}
  {Phys. Rev. Mater.}\ }\textbf {\bibinfo {volume} {3}},\ \bibinfo {pages}
  {034004} (\bibinfo {year} {2019})}\BibitemShut {NoStop}%
\bibitem [{\citenamefont {Rhyee}\ \emph {et~al.}(2009)\citenamefont {Rhyee},
  \citenamefont {Lee}, \citenamefont {Lee}, \citenamefont {Cho}, \citenamefont
  {Kim}, \citenamefont {Lee}, \citenamefont {Kwon}, \citenamefont {Shim},\ and\
  \citenamefont {Kotliar}}]{rhyeeN09}%
  \BibitemOpen
  \bibfield  {author} {\bibinfo {author} {\bibfnamefont {J.-S.}\ \bibnamefont
  {Rhyee}}, \bibinfo {author} {\bibfnamefont {K.~H.}\ \bibnamefont {Lee}},
  \bibinfo {author} {\bibfnamefont {S.~M.}\ \bibnamefont {Lee}}, \bibinfo
  {author} {\bibfnamefont {E.}~\bibnamefont {Cho}}, \bibinfo {author}
  {\bibfnamefont {S.~I.}\ \bibnamefont {Kim}}, \bibinfo {author} {\bibfnamefont
  {E.}~\bibnamefont {Lee}}, \bibinfo {author} {\bibfnamefont {Y.~S.}\
  \bibnamefont {Kwon}}, \bibinfo {author} {\bibfnamefont {J.~H.}\ \bibnamefont
  {Shim}}, \ and\ \bibinfo {author} {\bibfnamefont {G.}~\bibnamefont
  {Kotliar}},\ }\href {\doibase 10.1038/nature08088} {\bibfield  {journal}
  {\bibinfo  {journal} {Nature}\ }\textbf {\bibinfo {volume} {459}},\ \bibinfo
  {pages} {965} (\bibinfo {year} {2009})}\BibitemShut {NoStop}%
\bibitem [{\citenamefont {Han}\ \emph {et~al.}(2014{\natexlab{a}})\citenamefont
  {Han}, \citenamefont {Chen}, \citenamefont {Sun}, \citenamefont {Yang},
  \citenamefont {Cheng}, \citenamefont {Li}, \citenamefont {Lu}, \citenamefont
  {Gibbs}, \citenamefont {Snyder}, \citenamefont {Jack}, \citenamefont
  {Drennan},\ and\ \citenamefont {Zou}}]{hanCEC14}%
  \BibitemOpen
  \bibfield  {author} {\bibinfo {author} {\bibfnamefont {G.}~\bibnamefont
  {Han}}, \bibinfo {author} {\bibfnamefont {Z.-G.}\ \bibnamefont {Chen}},
  \bibinfo {author} {\bibfnamefont {C.}~\bibnamefont {Sun}}, \bibinfo {author}
  {\bibfnamefont {L.}~\bibnamefont {Yang}}, \bibinfo {author} {\bibfnamefont
  {L.}~\bibnamefont {Cheng}}, \bibinfo {author} {\bibfnamefont
  {Z.}~\bibnamefont {Li}}, \bibinfo {author} {\bibfnamefont {W.}~\bibnamefont
  {Lu}}, \bibinfo {author} {\bibfnamefont {Z.~M.}\ \bibnamefont {Gibbs}},
  \bibinfo {author} {\bibfnamefont {G.~J.}\ \bibnamefont {Snyder}}, \bibinfo
  {author} {\bibfnamefont {K.}~\bibnamefont {Jack}}, \bibinfo {author}
  {\bibfnamefont {J.}~\bibnamefont {Drennan}}, \ and\ \bibinfo {author}
  {\bibfnamefont {J.}~\bibnamefont {Zou}},\ }\href {\doibase
  10.1039/C3CE41815D} {\bibfield  {journal} {\bibinfo  {journal}
  {CrystEngComm}\ }\textbf {\bibinfo {volume} {16}},\ \bibinfo {pages} {393}
  (\bibinfo {year} {2014}{\natexlab{a}})}\BibitemShut {NoStop}%
\bibitem [{\citenamefont {Debbichi}\ \emph {et~al.}(2014)\citenamefont
  {Debbichi}, \citenamefont {Eriksson},\ and\ \citenamefont
  {Leb{\`e}gue}}]{debbichiAP14}%
  \BibitemOpen
  \bibfield  {author} {\bibinfo {author} {\bibfnamefont {L.}~\bibnamefont
  {Debbichi}}, \bibinfo {author} {\bibfnamefont {O.}~\bibnamefont {Eriksson}},
  \ and\ \bibinfo {author} {\bibfnamefont {S.}~\bibnamefont {Leb{\`e}gue}},\
  }\href {\doibase 10.1002/andp.201400159} {\bibfield  {journal} {\bibinfo
  {journal} {Ann. Phys.}\ }\textbf {\bibinfo {volume} {526}},\ \bibinfo {pages}
  {402} (\bibinfo {year} {2014})}\BibitemShut {NoStop}%
\bibitem [{\citenamefont {Bandurin}\ \emph {et~al.}(2016)\citenamefont
  {Bandurin}, \citenamefont {Tyurnina}, \citenamefont {Yu}, \citenamefont
  {Mishchenko}, \citenamefont {Z{\'o}lyomi}, \citenamefont {Morozov},
  \citenamefont {Kumar}, \citenamefont {Gorbachev}, \citenamefont {Kudrynskyi},
  \citenamefont {Pezzini}, \citenamefont {Kovalyuk}, \citenamefont {Zeitler},
  \citenamefont {Novoselov}, \citenamefont {Patan{\'e}}, \citenamefont {Eaves},
  \citenamefont {Grigorieva}, \citenamefont {Fal'ko}, \citenamefont {Geim},\
  and\ \citenamefont {Cao}}]{bandurinNN16}%
  \BibitemOpen
  \bibfield  {author} {\bibinfo {author} {\bibfnamefont {D.~A.}\ \bibnamefont
  {Bandurin}}, \bibinfo {author} {\bibfnamefont {A.~V.}\ \bibnamefont
  {Tyurnina}}, \bibinfo {author} {\bibfnamefont {G.~L.}\ \bibnamefont {Yu}},
  \bibinfo {author} {\bibfnamefont {A.}~\bibnamefont {Mishchenko}}, \bibinfo
  {author} {\bibfnamefont {V.}~\bibnamefont {Z{\'o}lyomi}}, \bibinfo {author}
  {\bibfnamefont {S.~V.}\ \bibnamefont {Morozov}}, \bibinfo {author}
  {\bibfnamefont {R.~K.}\ \bibnamefont {Kumar}}, \bibinfo {author}
  {\bibfnamefont {R.~V.}\ \bibnamefont {Gorbachev}}, \bibinfo {author}
  {\bibfnamefont {Z.~R.}\ \bibnamefont {Kudrynskyi}}, \bibinfo {author}
  {\bibfnamefont {S.}~\bibnamefont {Pezzini}}, \bibinfo {author} {\bibfnamefont
  {Z.~D.}\ \bibnamefont {Kovalyuk}}, \bibinfo {author} {\bibfnamefont
  {U.}~\bibnamefont {Zeitler}}, \bibinfo {author} {\bibfnamefont {K.~S.}\
  \bibnamefont {Novoselov}}, \bibinfo {author} {\bibfnamefont {A.}~\bibnamefont
  {Patan{\'e}}}, \bibinfo {author} {\bibfnamefont {L.}~\bibnamefont {Eaves}},
  \bibinfo {author} {\bibfnamefont {I.~V.}\ \bibnamefont {Grigorieva}},
  \bibinfo {author} {\bibfnamefont {V.~I.}\ \bibnamefont {Fal'ko}}, \bibinfo
  {author} {\bibfnamefont {A.~K.}\ \bibnamefont {Geim}}, \ and\ \bibinfo
  {author} {\bibfnamefont {Y.}~\bibnamefont {Cao}},\ }\href {\doibase
  10.1038/nnano.2016.242} {\bibfield  {journal} {\bibinfo  {journal} {Nat.
  Nanotechnol.}\ }\textbf {\bibinfo {volume} {12}},\ \bibinfo {pages} {223}
  (\bibinfo {year} {2016})}\BibitemShut {NoStop}%
\bibitem [{\citenamefont {Liu}\ \emph {et~al.}(2018)\citenamefont {Liu},
  \citenamefont {Chen},\ and\ \citenamefont {Li}}]{liuJACE18}%
  \BibitemOpen
  \bibfield  {author} {\bibinfo {author} {\bibfnamefont {G.}~\bibnamefont
  {Liu}}, \bibinfo {author} {\bibfnamefont {K.}~\bibnamefont {Chen}}, \ and\
  \bibinfo {author} {\bibfnamefont {J.}~\bibnamefont {Li}},\ }\href {\doibase
  10.1111/jace.15223} {\bibfield  {journal} {\bibinfo  {journal} {J. Am. Ceram.
  Soc.}\ }\textbf {\bibinfo {volume} {101}},\ \bibinfo {pages} {36} (\bibinfo
  {year} {2018})}\BibitemShut {NoStop}%
\bibitem [{\citenamefont {Ding}\ \emph {et~al.}(2017)\citenamefont {Ding},
  \citenamefont {Zhu}, \citenamefont {Wang}, \citenamefont {Gao}, \citenamefont
  {Xiao}, \citenamefont {Gu}, \citenamefont {Zhang},\ and\ \citenamefont
  {Zhu}}]{dingNC17}%
  \BibitemOpen
  \bibfield  {author} {\bibinfo {author} {\bibfnamefont {W.}~\bibnamefont
  {Ding}}, \bibinfo {author} {\bibfnamefont {J.}~\bibnamefont {Zhu}}, \bibinfo
  {author} {\bibfnamefont {Z.}~\bibnamefont {Wang}}, \bibinfo {author}
  {\bibfnamefont {Y.}~\bibnamefont {Gao}}, \bibinfo {author} {\bibfnamefont
  {D.}~\bibnamefont {Xiao}}, \bibinfo {author} {\bibfnamefont {Y.}~\bibnamefont
  {Gu}}, \bibinfo {author} {\bibfnamefont {Z.}~\bibnamefont {Zhang}}, \ and\
  \bibinfo {author} {\bibfnamefont {W.}~\bibnamefont {Zhu}},\ }\href {\doibase
  10.1038/ncomms14956} {\bibfield  {journal} {\bibinfo  {journal} {Nat.
  Commun.}\ }\textbf {\bibinfo {volume} {8}},\ \bibinfo {pages} {14956}
  (\bibinfo {year} {2017})}\BibitemShut {NoStop}%
\bibitem [{\citenamefont {Scott}(2007)}]{scottSCI07}%
  \BibitemOpen
  \bibfield  {author} {\bibinfo {author} {\bibfnamefont {J.~F.}\ \bibnamefont
  {Scott}},\ }\href {\doibase 10.1126/science.1129564} {\bibfield  {journal}
  {\bibinfo  {journal} {Science}\ }\textbf {\bibinfo {volume} {315}},\ \bibinfo
  {pages} {954} (\bibinfo {year} {2007})}\BibitemShut {NoStop}%
\bibitem [{\citenamefont {Miller}\ and\ \citenamefont
  {McWhorter}(1992)}]{millerJAP92}%
  \BibitemOpen
  \bibfield  {author} {\bibinfo {author} {\bibfnamefont {S.~L.}\ \bibnamefont
  {Miller}}\ and\ \bibinfo {author} {\bibfnamefont {P.~J.}\ \bibnamefont
  {McWhorter}},\ }\href {\doibase 10.1063/1.351910} {\bibfield  {journal}
  {\bibinfo  {journal} {J. Appl. Phys.}\ }\textbf {\bibinfo {volume} {72}},\
  \bibinfo {pages} {5999} (\bibinfo {year} {1992})}\BibitemShut {NoStop}%
\bibitem [{\citenamefont {Park}\ \emph {et~al.}(1999)\citenamefont {Park},
  \citenamefont {Kang}, \citenamefont {Bu}, \citenamefont {Noh}, \citenamefont
  {Lee},\ and\ \citenamefont {Jo}}]{parkNat99}%
  \BibitemOpen
  \bibfield  {author} {\bibinfo {author} {\bibfnamefont {B.~H.}\ \bibnamefont
  {Park}}, \bibinfo {author} {\bibfnamefont {B.~S.}\ \bibnamefont {Kang}},
  \bibinfo {author} {\bibfnamefont {S.~D.}\ \bibnamefont {Bu}}, \bibinfo
  {author} {\bibfnamefont {T.~W.}\ \bibnamefont {Noh}}, \bibinfo {author}
  {\bibfnamefont {J.}~\bibnamefont {Lee}}, \ and\ \bibinfo {author}
  {\bibfnamefont {W.}~\bibnamefont {Jo}},\ }\href {\doibase 10.1038/44352}
  {\bibfield  {journal} {\bibinfo  {journal} {Nature}\ }\textbf {\bibinfo
  {volume} {401}},\ \bibinfo {pages} {682} (\bibinfo {year}
  {1999})}\BibitemShut {NoStop}%
\bibitem [{\citenamefont {Wan}\ \emph {et~al.}(2018)\citenamefont {Wan},
  \citenamefont {Li}, \citenamefont {Li}, \citenamefont {Mao}, \citenamefont
  {Zhu},\ and\ \citenamefont {Zeng}}]{wanNS18}%
  \BibitemOpen
  \bibfield  {author} {\bibinfo {author} {\bibfnamefont {S.}~\bibnamefont
  {Wan}}, \bibinfo {author} {\bibfnamefont {Y.}~\bibnamefont {Li}}, \bibinfo
  {author} {\bibfnamefont {W.}~\bibnamefont {Li}}, \bibinfo {author}
  {\bibfnamefont {X.}~\bibnamefont {Mao}}, \bibinfo {author} {\bibfnamefont
  {W.}~\bibnamefont {Zhu}}, \ and\ \bibinfo {author} {\bibfnamefont
  {H.}~\bibnamefont {Zeng}},\ }\href {\doibase 10.1039/C8NR04422H} {\bibfield
  {journal} {\bibinfo  {journal} {Nanoscale}\ }\textbf {\bibinfo {volume}
  {10}},\ \bibinfo {pages} {14885} (\bibinfo {year} {2018})}\BibitemShut
  {NoStop}%
\bibitem [{\citenamefont {Ponath}\ \emph {et~al.}(2015)\citenamefont {Ponath},
  \citenamefont {Fredrickson}, \citenamefont {Posadas}, \citenamefont {Ren},
  \citenamefont {Wu}, \citenamefont {Vasudevan}, \citenamefont {Okatan},
  \citenamefont {Jesse}, \citenamefont {Aoki}, \citenamefont {McCartney},
  \citenamefont {Smith}, \citenamefont {Kalinin}, \citenamefont {Lai},\ and\
  \citenamefont {Demkov}}]{ponathNC15}%
  \BibitemOpen
  \bibfield  {author} {\bibinfo {author} {\bibfnamefont {P.}~\bibnamefont
  {Ponath}}, \bibinfo {author} {\bibfnamefont {K.}~\bibnamefont {Fredrickson}},
  \bibinfo {author} {\bibfnamefont {A.~B.}\ \bibnamefont {Posadas}}, \bibinfo
  {author} {\bibfnamefont {Y.}~\bibnamefont {Ren}}, \bibinfo {author}
  {\bibfnamefont {X.}~\bibnamefont {Wu}}, \bibinfo {author} {\bibfnamefont
  {R.~K.}\ \bibnamefont {Vasudevan}}, \bibinfo {author} {\bibfnamefont {M.~B.}\
  \bibnamefont {Okatan}}, \bibinfo {author} {\bibfnamefont {S.}~\bibnamefont
  {Jesse}}, \bibinfo {author} {\bibfnamefont {T.}~\bibnamefont {Aoki}},
  \bibinfo {author} {\bibfnamefont {M.~R.}\ \bibnamefont {McCartney}}, \bibinfo
  {author} {\bibfnamefont {D.~J.}\ \bibnamefont {Smith}}, \bibinfo {author}
  {\bibfnamefont {S.~V.}\ \bibnamefont {Kalinin}}, \bibinfo {author}
  {\bibfnamefont {K.}~\bibnamefont {Lai}}, \ and\ \bibinfo {author}
  {\bibfnamefont {A.~A.}\ \bibnamefont {Demkov}},\ }\href {\doibase
  10.1038/ncomms7067} {\bibfield  {journal} {\bibinfo  {journal} {Nat.
  Commun.}\ }\textbf {\bibinfo {volume} {6}},\ \bibinfo {pages} {6067}
  (\bibinfo {year} {2015})}\BibitemShut {NoStop}%
\bibitem [{\citenamefont {Cho}\ \emph {et~al.}(2017)\citenamefont {Cho},
  \citenamefont {Ponath}, \citenamefont {Zheng}, \citenamefont {Hatanpaa},
  \citenamefont {Lai}, \citenamefont {Demkov},\ and\ \citenamefont
  {Downer}}]{choAPL18}%
  \BibitemOpen
  \bibfield  {author} {\bibinfo {author} {\bibfnamefont {Y.}~\bibnamefont
  {Cho}}, \bibinfo {author} {\bibfnamefont {P.}~\bibnamefont {Ponath}},
  \bibinfo {author} {\bibfnamefont {L.}~\bibnamefont {Zheng}}, \bibinfo
  {author} {\bibfnamefont {B.}~\bibnamefont {Hatanpaa}}, \bibinfo {author}
  {\bibfnamefont {K.}~\bibnamefont {Lai}}, \bibinfo {author} {\bibfnamefont
  {A.~A.}\ \bibnamefont {Demkov}}, \ and\ \bibinfo {author} {\bibfnamefont
  {M.~C.}\ \bibnamefont {Downer}},\ }\href {\doibase 10.1063/1.5020549}
  {\bibfield  {journal} {\bibinfo  {journal} {Appl. Phys. Lett.}\ }\textbf
  {\bibinfo {volume} {112}},\ \bibinfo {pages} {162901} (\bibinfo {year}
  {2017})}\BibitemShut {NoStop}%
\bibitem [{\citenamefont {Fong}\ \emph {et~al.}(2004)\citenamefont {Fong},
  \citenamefont {Stephenson}, \citenamefont {Streiffer}, \citenamefont
  {Eastman}, \citenamefont {Auciello}, \citenamefont {Fuoss},\ and\
  \citenamefont {Thompson}}]{fongSCI04}%
  \BibitemOpen
  \bibfield  {author} {\bibinfo {author} {\bibfnamefont {D.~D.}\ \bibnamefont
  {Fong}}, \bibinfo {author} {\bibfnamefont {G.~B.}\ \bibnamefont
  {Stephenson}}, \bibinfo {author} {\bibfnamefont {S.~K.}\ \bibnamefont
  {Streiffer}}, \bibinfo {author} {\bibfnamefont {J.~A.}\ \bibnamefont
  {Eastman}}, \bibinfo {author} {\bibfnamefont {O.}~\bibnamefont {Auciello}},
  \bibinfo {author} {\bibfnamefont {P.~H.}\ \bibnamefont {Fuoss}}, \ and\
  \bibinfo {author} {\bibfnamefont {C.}~\bibnamefont {Thompson}},\ }\href
  {\doibase 10.1126/science.1098252} {\bibfield  {journal} {\bibinfo  {journal}
  {Science}\ }\textbf {\bibinfo {volume} {304}},\ \bibinfo {pages} {1650}
  (\bibinfo {year} {2004})}\BibitemShut {NoStop}%
\bibitem [{\citenamefont {Meyer}\ and\ \citenamefont
  {Vanderbilt}(2001)}]{meyerPRB01}%
  \BibitemOpen
  \bibfield  {author} {\bibinfo {author} {\bibfnamefont {B.}~\bibnamefont
  {Meyer}}\ and\ \bibinfo {author} {\bibfnamefont {D.}~\bibnamefont
  {Vanderbilt}},\ }\href {\doibase 10.1103/PhysRevB.63.205426} {\bibfield
  {journal} {\bibinfo  {journal} {Phys. Rev. B}\ }\textbf {\bibinfo {volume}
  {63}},\ \bibinfo {pages} {205426} (\bibinfo {year} {2001})}\BibitemShut
  {NoStop}%
\bibitem [{\citenamefont {Han}\ \emph {et~al.}(2014{\natexlab{b}})\citenamefont
  {Han}, \citenamefont {Marshall}, \citenamefont {Wu}, \citenamefont
  {Schofield}, \citenamefont {Aoki}, \citenamefont {Twesten}, \citenamefont
  {Hoffman}, \citenamefont {Walker}, \citenamefont {Ahn},\ and\ \citenamefont
  {Y.}}]{hanNC14}%
  \BibitemOpen
  \bibfield  {author} {\bibinfo {author} {\bibfnamefont {M.-G.}\ \bibnamefont
  {Han}}, \bibinfo {author} {\bibfnamefont {M.~S.~J.}\ \bibnamefont
  {Marshall}}, \bibinfo {author} {\bibfnamefont {L.}~\bibnamefont {Wu}},
  \bibinfo {author} {\bibfnamefont {M.~A.}\ \bibnamefont {Schofield}}, \bibinfo
  {author} {\bibfnamefont {T.}~\bibnamefont {Aoki}}, \bibinfo {author}
  {\bibfnamefont {R.}~\bibnamefont {Twesten}}, \bibinfo {author} {\bibfnamefont
  {J.}~\bibnamefont {Hoffman}}, \bibinfo {author} {\bibfnamefont {F.~J.}\
  \bibnamefont {Walker}}, \bibinfo {author} {\bibfnamefont {C.~H.}\
  \bibnamefont {Ahn}}, \ and\ \bibinfo {author} {\bibfnamefont
  {Z.}~\bibnamefont {Y.}},\ }\href {\doibase 10.1038/ncomms5693} {\bibfield
  {journal} {\bibinfo  {journal} {Nat. Commun.}\ }\textbf {\bibinfo {volume}
  {5}},\ \bibinfo {pages} {4693} (\bibinfo {year}
  {2014}{\natexlab{b}})}\BibitemShut {NoStop}%
\bibitem [{\citenamefont {Dawber}\ \emph {et~al.}(2005)\citenamefont {Dawber},
  \citenamefont {Rabe},\ and\ \citenamefont {Scott}}]{dawberRMP05}%
  \BibitemOpen
  \bibfield  {author} {\bibinfo {author} {\bibfnamefont {M.}~\bibnamefont
  {Dawber}}, \bibinfo {author} {\bibfnamefont {K.~M.}\ \bibnamefont {Rabe}}, \
  and\ \bibinfo {author} {\bibfnamefont {J.~F.}\ \bibnamefont {Scott}},\ }\href
  {\doibase 10.1103/RevModPhys.77.1083} {\bibfield  {journal} {\bibinfo
  {journal} {Rev. Mod. Phys.}\ }\textbf {\bibinfo {volume} {77}},\ \bibinfo
  {pages} {1083} (\bibinfo {year} {2005})}\BibitemShut {NoStop}%
\bibitem [{\citenamefont {Chang}\ \emph {et~al.}(2016)\citenamefont {Chang},
  \citenamefont {Liu}, \citenamefont {Lin}, \citenamefont {Wang}, \citenamefont
  {Zhao}, \citenamefont {Zhang}, \citenamefont {Jin}, \citenamefont {Zhong},
  \citenamefont {Hu}, \citenamefont {Duan}, \citenamefont {Zhang},
  \citenamefont {Fu}, \citenamefont {Xue}, \citenamefont {Chen},\ and\
  \citenamefont {Ji}}]{changSCI16}%
  \BibitemOpen
  \bibfield  {author} {\bibinfo {author} {\bibfnamefont {K.}~\bibnamefont
  {Chang}}, \bibinfo {author} {\bibfnamefont {J.}~\bibnamefont {Liu}}, \bibinfo
  {author} {\bibfnamefont {H.}~\bibnamefont {Lin}}, \bibinfo {author}
  {\bibfnamefont {N.}~\bibnamefont {Wang}}, \bibinfo {author} {\bibfnamefont
  {K.}~\bibnamefont {Zhao}}, \bibinfo {author} {\bibfnamefont {A.}~\bibnamefont
  {Zhang}}, \bibinfo {author} {\bibfnamefont {F.}~\bibnamefont {Jin}}, \bibinfo
  {author} {\bibfnamefont {Y.}~\bibnamefont {Zhong}}, \bibinfo {author}
  {\bibfnamefont {X.}~\bibnamefont {Hu}}, \bibinfo {author} {\bibfnamefont
  {W.}~\bibnamefont {Duan}}, \bibinfo {author} {\bibfnamefont {Q.}~\bibnamefont
  {Zhang}}, \bibinfo {author} {\bibfnamefont {L.}~\bibnamefont {Fu}}, \bibinfo
  {author} {\bibfnamefont {Q.-K.}\ \bibnamefont {Xue}}, \bibinfo {author}
  {\bibfnamefont {X.}~\bibnamefont {Chen}}, \ and\ \bibinfo {author}
  {\bibfnamefont {S.-H.}\ \bibnamefont {Ji}},\ }\href {\doibase
  10.1126/science.aad8609} {\bibfield  {journal} {\bibinfo  {journal}
  {Science}\ }\textbf {\bibinfo {volume} {353}},\ \bibinfo {pages} {274}
  (\bibinfo {year} {2016})}\BibitemShut {NoStop}%
\bibitem [{\citenamefont {Shirodkar}\ and\ \citenamefont
  {Waghmare}(2014)}]{shirodkarPRL14}%
  \BibitemOpen
  \bibfield  {author} {\bibinfo {author} {\bibfnamefont {S.~N.}\ \bibnamefont
  {Shirodkar}}\ and\ \bibinfo {author} {\bibfnamefont {U.~V.}\ \bibnamefont
  {Waghmare}},\ }\href {\doibase 10.1103/PhysRevLett.112.157601} {\bibfield
  {journal} {\bibinfo  {journal} {Phys. Rev. Lett.}\ }\textbf {\bibinfo
  {volume} {112}},\ \bibinfo {pages} {157601} (\bibinfo {year}
  {2014})}\BibitemShut {NoStop}%
\bibitem [{\citenamefont {Wu}\ and\ \citenamefont {Zeng}(2016)}]{wuNL16}%
  \BibitemOpen
  \bibfield  {author} {\bibinfo {author} {\bibfnamefont {M.}~\bibnamefont
  {Wu}}\ and\ \bibinfo {author} {\bibfnamefont {X.~C.}\ \bibnamefont {Zeng}},\
  }\href {\doibase 10.1021/acs.nanolett.6b00726} {\bibfield  {journal}
  {\bibinfo  {journal} {Nano Lett.}\ }\textbf {\bibinfo {volume} {16}},\
  \bibinfo {pages} {3236} (\bibinfo {year} {2016})}\BibitemShut {NoStop}%
\bibitem [{\citenamefont {Li}\ \emph {et~al.}(2018)\citenamefont {Li},
  \citenamefont {Sabino}, \citenamefont {Crasto~de Lima}, \citenamefont {Wang},
  \citenamefont {Miwa},\ and\ \citenamefont {Janotti}}]{liPRB18}%
  \BibitemOpen
  \bibfield  {author} {\bibinfo {author} {\bibfnamefont {W.}~\bibnamefont
  {Li}}, \bibinfo {author} {\bibfnamefont {F.~P.}\ \bibnamefont {Sabino}},
  \bibinfo {author} {\bibfnamefont {F.}~\bibnamefont {Crasto~de Lima}},
  \bibinfo {author} {\bibfnamefont {T.}~\bibnamefont {Wang}}, \bibinfo {author}
  {\bibfnamefont {R.~H.}\ \bibnamefont {Miwa}}, \ and\ \bibinfo {author}
  {\bibfnamefont {A.}~\bibnamefont {Janotti}},\ }\href {\doibase
  10.1103/PhysRevB.98.165134} {\bibfield  {journal} {\bibinfo  {journal} {Phys.
  Rev. B}\ }\textbf {\bibinfo {volume} {98}},\ \bibinfo {pages} {165134}
  (\bibinfo {year} {2018})}\BibitemShut {NoStop}%
\bibitem [{\citenamefont {Wan}\ \emph {et~al.}(2019)\citenamefont {Wan},
  \citenamefont {Li}, \citenamefont {Li}, \citenamefont {Mao}, \citenamefont
  {Wang}, \citenamefont {Chen}, \citenamefont {Dong}, \citenamefont {Nie},
  \citenamefont {Xiang}, \citenamefont {Liu}, \citenamefont {Zhu},\ and\
  \citenamefont {Zeng}}]{wanAFM19}%
  \BibitemOpen
  \bibfield  {author} {\bibinfo {author} {\bibfnamefont {S.}~\bibnamefont
  {Wan}}, \bibinfo {author} {\bibfnamefont {Y.}~\bibnamefont {Li}}, \bibinfo
  {author} {\bibfnamefont {W.}~\bibnamefont {Li}}, \bibinfo {author}
  {\bibfnamefont {X.}~\bibnamefont {Mao}}, \bibinfo {author} {\bibfnamefont
  {C.}~\bibnamefont {Wang}}, \bibinfo {author} {\bibfnamefont {C.}~\bibnamefont
  {Chen}}, \bibinfo {author} {\bibfnamefont {J.}~\bibnamefont {Dong}}, \bibinfo
  {author} {\bibfnamefont {A.}~\bibnamefont {Nie}}, \bibinfo {author}
  {\bibfnamefont {J.}~\bibnamefont {Xiang}}, \bibinfo {author} {\bibfnamefont
  {Z.}~\bibnamefont {Liu}}, \bibinfo {author} {\bibfnamefont {W.}~\bibnamefont
  {Zhu}}, \ and\ \bibinfo {author} {\bibfnamefont {H.}~\bibnamefont {Zeng}},\
  }\href {\doibase 10.1002/adfm.201808606} {\bibfield  {journal} {\bibinfo
  {journal} {Adv. Funct. Mater.}\ ,\ \bibinfo {pages} {1808606}} (\bibinfo
  {year} {2019})}\BibitemShut {NoStop}%
\bibitem [{\citenamefont {Yang}\ and\ \citenamefont {Hao}(2019)}]{yangAMT19}%
  \BibitemOpen
  \bibfield  {author} {\bibinfo {author} {\bibfnamefont {Z.}~\bibnamefont
  {Yang}}\ and\ \bibinfo {author} {\bibfnamefont {J.}~\bibnamefont {Hao}},\
  }\href {\doibase 10.1002/admt.201900108} {\bibfield  {journal} {\bibinfo
  {journal} {Adv. Mater. Technol.}\ }\textbf {\bibinfo {volume} {4}},\ \bibinfo
  {pages} {1900108} (\bibinfo {year} {2019})}\BibitemShut {NoStop}%
\bibitem [{\citenamefont {Quereda}\ \emph {et~al.}(2016)\citenamefont
  {Quereda}, \citenamefont {Biele}, \citenamefont {Rubio-Bollinger},
  \citenamefont {Agra{\"i}t}, \citenamefont {D'Agosta},\ and\ \citenamefont
  {Castellanos-Gomez}}]{queredaAOM16}%
  \BibitemOpen
  \bibfield  {author} {\bibinfo {author} {\bibfnamefont {J.}~\bibnamefont
  {Quereda}}, \bibinfo {author} {\bibfnamefont {R.}~\bibnamefont {Biele}},
  \bibinfo {author} {\bibfnamefont {G.}~\bibnamefont {Rubio-Bollinger}},
  \bibinfo {author} {\bibfnamefont {N.}~\bibnamefont {Agra{\"i}t}}, \bibinfo
  {author} {\bibfnamefont {R.}~\bibnamefont {D'Agosta}}, \ and\ \bibinfo
  {author} {\bibfnamefont {A.}~\bibnamefont {Castellanos-Gomez}},\ }\href
  {\doibase 10.1002/adom.201600365} {\bibfield  {journal} {\bibinfo  {journal}
  {Adv. Opt. Mater}\ }\textbf {\bibinfo {volume} {4}},\ \bibinfo {pages} {1939}
  (\bibinfo {year} {2016})}\BibitemShut {NoStop}%
\bibitem [{\citenamefont {Wu}\ \emph {et~al.}(2015)\citenamefont {Wu},
  \citenamefont {Pak}, \citenamefont {Liu}, \citenamefont {Zhou}, \citenamefont
  {Wu}, \citenamefont {Zhu}, \citenamefont {Lin}, \citenamefont {Han},
  \citenamefont {Ren}, \citenamefont {Peng}, \citenamefont {Tsai},
  \citenamefont {Hwang},\ and\ \citenamefont {Lai}}]{wuNL15}%
  \BibitemOpen
  \bibfield  {author} {\bibinfo {author} {\bibfnamefont {D.}~\bibnamefont
  {Wu}}, \bibinfo {author} {\bibfnamefont {A.~J.}\ \bibnamefont {Pak}},
  \bibinfo {author} {\bibfnamefont {Y.}~\bibnamefont {Liu}}, \bibinfo {author}
  {\bibfnamefont {Y.}~\bibnamefont {Zhou}}, \bibinfo {author} {\bibfnamefont
  {X.}~\bibnamefont {Wu}}, \bibinfo {author} {\bibfnamefont {Y.}~\bibnamefont
  {Zhu}}, \bibinfo {author} {\bibfnamefont {M.}~\bibnamefont {Lin}}, \bibinfo
  {author} {\bibfnamefont {Y.}~\bibnamefont {Han}}, \bibinfo {author}
  {\bibfnamefont {Y.}~\bibnamefont {Ren}}, \bibinfo {author} {\bibfnamefont
  {H.}~\bibnamefont {Peng}}, \bibinfo {author} {\bibfnamefont {Y.-H.}\
  \bibnamefont {Tsai}}, \bibinfo {author} {\bibfnamefont {G.~S.}\ \bibnamefont
  {Hwang}}, \ and\ \bibinfo {author} {\bibfnamefont {K.}~\bibnamefont {Lai}},\
  }\href {\doibase 10.1021/acs.nanolett.5b03575} {\bibfield  {journal}
  {\bibinfo  {journal} {Nano Lett.}\ }\textbf {\bibinfo {volume} {15}},\
  \bibinfo {pages} {8136} (\bibinfo {year} {2015})}\BibitemShut {NoStop}%
\bibitem [{\citenamefont {Hu}\ and\ \citenamefont {Huang}(2017)}]{huRA17}%
  \BibitemOpen
  \bibfield  {author} {\bibinfo {author} {\bibfnamefont {L.}~\bibnamefont
  {Hu}}\ and\ \bibinfo {author} {\bibfnamefont {X.}~\bibnamefont {Huang}},\
  }\href {\doibase 10.1039/C7RA11014F} {\bibfield  {journal} {\bibinfo
  {journal} {RSC Adv.}\ }\textbf {\bibinfo {volume} {7}},\ \bibinfo {pages}
  {55034} (\bibinfo {year} {2017})}\BibitemShut {NoStop}%
\bibitem [{\citenamefont {Tao}\ and\ \citenamefont {Gu}(2013)}]{taoNL13}%
  \BibitemOpen
  \bibfield  {author} {\bibinfo {author} {\bibfnamefont {X.}~\bibnamefont
  {Tao}}\ and\ \bibinfo {author} {\bibfnamefont {Y.}~\bibnamefont {Gu}},\
  }\href {\doibase 10.1021/nl400888p} {\bibfield  {journal} {\bibinfo
  {journal} {Nano Lett.}\ }\textbf {\bibinfo {volume} {13}},\ \bibinfo {pages}
  {3501} (\bibinfo {year} {2013})}\BibitemShut {NoStop}%
\bibitem [{\citenamefont {Yu}\ \emph {et~al.}(2007)\citenamefont {Yu},
  \citenamefont {Ju}, \citenamefont {Sun}, \citenamefont {Ng}, \citenamefont
  {Nguyen}, \citenamefont {Meyyappan},\ and\ \citenamefont {Janes}}]{yuAPL07}%
  \BibitemOpen
  \bibfield  {author} {\bibinfo {author} {\bibfnamefont {B.}~\bibnamefont
  {Yu}}, \bibinfo {author} {\bibfnamefont {S.}~\bibnamefont {Ju}}, \bibinfo
  {author} {\bibfnamefont {X.}~\bibnamefont {Sun}}, \bibinfo {author}
  {\bibfnamefont {G.}~\bibnamefont {Ng}}, \bibinfo {author} {\bibfnamefont
  {T.~D.}\ \bibnamefont {Nguyen}}, \bibinfo {author} {\bibfnamefont
  {M.}~\bibnamefont {Meyyappan}}, \ and\ \bibinfo {author} {\bibfnamefont
  {D.~B.}\ \bibnamefont {Janes}},\ }\href {\doibase 10.1063/1.2793505}
  {\bibfield  {journal} {\bibinfo  {journal} {Appl. Phys. Lett.}\ }\textbf
  {\bibinfo {volume} {91}},\ \bibinfo {pages} {133119} (\bibinfo {year}
  {2007})}\BibitemShut {NoStop}%
\bibitem [{\citenamefont {Choi}\ \emph {et~al.}(2017)\citenamefont {Choi},
  \citenamefont {Cheong}, \citenamefont {Ra}, \citenamefont {Lee},
  \citenamefont {Bae}, \citenamefont {Lee}, \citenamefont {Lee}, \citenamefont
  {Yang}, \citenamefont {Hone},\ and\ \citenamefont {Yoo}}]{choiAM17}%
  \BibitemOpen
  \bibfield  {author} {\bibinfo {author} {\bibfnamefont {M.~S.}\ \bibnamefont
  {Choi}}, \bibinfo {author} {\bibfnamefont {B.}~\bibnamefont {Cheong}},
  \bibinfo {author} {\bibfnamefont {C.~H.}\ \bibnamefont {Ra}}, \bibinfo
  {author} {\bibfnamefont {S.}~\bibnamefont {Lee}}, \bibinfo {author}
  {\bibfnamefont {J.-H.}\ \bibnamefont {Bae}}, \bibinfo {author} {\bibfnamefont
  {S.}~\bibnamefont {Lee}}, \bibinfo {author} {\bibfnamefont {G.-D.}\
  \bibnamefont {Lee}}, \bibinfo {author} {\bibfnamefont {C.-W.}\ \bibnamefont
  {Yang}}, \bibinfo {author} {\bibfnamefont {J.}~\bibnamefont {Hone}}, \ and\
  \bibinfo {author} {\bibfnamefont {W.~J.}\ \bibnamefont {Yoo}},\ }\href
  {\doibase 10.1002/adma.201703568} {\bibfield  {journal} {\bibinfo  {journal}
  {Adv. Mater.}\ }\textbf {\bibinfo {volume} {29}},\ \bibinfo {pages} {1703568}
  (\bibinfo {year} {2017})}\BibitemShut {NoStop}%
\bibitem [{\citenamefont {Zhou}\ \emph {et~al.}(2017)\citenamefont {Zhou},
  \citenamefont {Wu}, \citenamefont {Zhu}, \citenamefont {Cho}, \citenamefont
  {He}, \citenamefont {Yang}, \citenamefont {Herrera}, \citenamefont {Chu},
  \citenamefont {Han}, \citenamefont {Downer}, \citenamefont {Peng},\ and\
  \citenamefont {Lai}}]{zhouNL17}%
  \BibitemOpen
  \bibfield  {author} {\bibinfo {author} {\bibfnamefont {Y.}~\bibnamefont
  {Zhou}}, \bibinfo {author} {\bibfnamefont {D.}~\bibnamefont {Wu}}, \bibinfo
  {author} {\bibfnamefont {Y.}~\bibnamefont {Zhu}}, \bibinfo {author}
  {\bibfnamefont {Y.}~\bibnamefont {Cho}}, \bibinfo {author} {\bibfnamefont
  {Q.}~\bibnamefont {He}}, \bibinfo {author} {\bibfnamefont {X.}~\bibnamefont
  {Yang}}, \bibinfo {author} {\bibfnamefont {K.}~\bibnamefont {Herrera}},
  \bibinfo {author} {\bibfnamefont {Z.}~\bibnamefont {Chu}}, \bibinfo {author}
  {\bibfnamefont {Y.}~\bibnamefont {Han}}, \bibinfo {author} {\bibfnamefont
  {M.~C.}\ \bibnamefont {Downer}}, \bibinfo {author} {\bibfnamefont
  {H.}~\bibnamefont {Peng}}, \ and\ \bibinfo {author} {\bibfnamefont
  {K.}~\bibnamefont {Lai}},\ }\href {\doibase 10.1021/acs.nanolett.7b02198}
  {\bibfield  {journal} {\bibinfo  {journal} {Nano Lett.}\ }\textbf {\bibinfo
  {volume} {17}},\ \bibinfo {pages} {5508} (\bibinfo {year}
  {2017})}\BibitemShut {NoStop}%
\bibitem [{\citenamefont {Xiao}\ \emph {et~al.}(2018)\citenamefont {Xiao},
  \citenamefont {Zhu}, \citenamefont {Wang}, \citenamefont {Feng},
  \citenamefont {Hu}, \citenamefont {Dasgupta}, \citenamefont {Han},
  \citenamefont {Wang}, \citenamefont {Muller}, \citenamefont {Martin},
  \citenamefont {Hu},\ and\ \citenamefont {Zhang}}]{xiaoPRL18}%
  \BibitemOpen
  \bibfield  {author} {\bibinfo {author} {\bibfnamefont {J.}~\bibnamefont
  {Xiao}}, \bibinfo {author} {\bibfnamefont {H.}~\bibnamefont {Zhu}}, \bibinfo
  {author} {\bibfnamefont {Y.}~\bibnamefont {Wang}}, \bibinfo {author}
  {\bibfnamefont {W.}~\bibnamefont {Feng}}, \bibinfo {author} {\bibfnamefont
  {Y.}~\bibnamefont {Hu}}, \bibinfo {author} {\bibfnamefont {A.}~\bibnamefont
  {Dasgupta}}, \bibinfo {author} {\bibfnamefont {Y.}~\bibnamefont {Han}},
  \bibinfo {author} {\bibfnamefont {Y.}~\bibnamefont {Wang}}, \bibinfo {author}
  {\bibfnamefont {D.~A.}\ \bibnamefont {Muller}}, \bibinfo {author}
  {\bibfnamefont {L.~W.}\ \bibnamefont {Martin}}, \bibinfo {author}
  {\bibfnamefont {P.}~\bibnamefont {Hu}}, \ and\ \bibinfo {author}
  {\bibfnamefont {X.}~\bibnamefont {Zhang}},\ }\href {\doibase
  10.1103/PhysRevLett.120.227601} {\bibfield  {journal} {\bibinfo  {journal}
  {Phys. Rev. Lett.}\ }\textbf {\bibinfo {volume} {120}} (\bibinfo {year}
  {2018}),\ 10.1103/PhysRevLett.120.227601}\BibitemShut {NoStop}%
\bibitem [{\citenamefont {Xue}\ \emph {et~al.}(2018)\citenamefont {Xue},
  \citenamefont {Hu}, \citenamefont {Lee}, \citenamefont {Lu}, \citenamefont
  {Zhang}, \citenamefont {Tang}, \citenamefont {Han}, \citenamefont {Hsu},
  \citenamefont {Tu}, \citenamefont {Chang}, \citenamefont {Lien},
  \citenamefont {He}, \citenamefont {Zhang}, \citenamefont {Li},\ and\
  \citenamefont {Zhang}}]{xueAFM18}%
  \BibitemOpen
  \bibfield  {author} {\bibinfo {author} {\bibfnamefont {F.}~\bibnamefont
  {Xue}}, \bibinfo {author} {\bibfnamefont {W.}~\bibnamefont {Hu}}, \bibinfo
  {author} {\bibfnamefont {K.-C.}\ \bibnamefont {Lee}}, \bibinfo {author}
  {\bibfnamefont {L.-S.}\ \bibnamefont {Lu}}, \bibinfo {author} {\bibfnamefont
  {J.}~\bibnamefont {Zhang}}, \bibinfo {author} {\bibfnamefont {H.-L.}\
  \bibnamefont {Tang}}, \bibinfo {author} {\bibfnamefont {A.}~\bibnamefont
  {Han}}, \bibinfo {author} {\bibfnamefont {W.-T.}\ \bibnamefont {Hsu}},
  \bibinfo {author} {\bibfnamefont {S.}~\bibnamefont {Tu}}, \bibinfo {author}
  {\bibfnamefont {W.-H.}\ \bibnamefont {Chang}}, \bibinfo {author}
  {\bibfnamefont {C.-H.}\ \bibnamefont {Lien}}, \bibinfo {author}
  {\bibfnamefont {J.-H.}\ \bibnamefont {He}}, \bibinfo {author} {\bibfnamefont
  {Z.}~\bibnamefont {Zhang}}, \bibinfo {author} {\bibfnamefont {L.-J.}\
  \bibnamefont {Li}}, \ and\ \bibinfo {author} {\bibfnamefont {X.}~\bibnamefont
  {Zhang}},\ }\href {\doibase 10.1002/adfm.201803738} {\bibfield  {journal}
  {\bibinfo  {journal} {Adv. Funct. Mater.}\ }\textbf {\bibinfo {volume}
  {28}},\ \bibinfo {pages} {1803738} (\bibinfo {year} {2018})}\BibitemShut
  {NoStop}%
\bibitem [{\citenamefont {Zheng}\ \emph {et~al.}(2018)\citenamefont {Zheng},
  \citenamefont {Yu}, \citenamefont {Zhu}, \citenamefont {Collins},
  \citenamefont {Kim}, \citenamefont {Lou}, \citenamefont {Xu}, \citenamefont
  {Li}, \citenamefont {Wei}, \citenamefont {Zhang}, \citenamefont {Edmonds},
  \citenamefont {Li}, \citenamefont {Seidel}, \citenamefont {Zhu},
  \citenamefont {Liu}, \citenamefont {Tang},\ and\ \citenamefont
  {Fuhrer}}]{zhengSA18}%
  \BibitemOpen
  \bibfield  {author} {\bibinfo {author} {\bibfnamefont {C.}~\bibnamefont
  {Zheng}}, \bibinfo {author} {\bibfnamefont {L.}~\bibnamefont {Yu}}, \bibinfo
  {author} {\bibfnamefont {L.}~\bibnamefont {Zhu}}, \bibinfo {author}
  {\bibfnamefont {J.~L.}\ \bibnamefont {Collins}}, \bibinfo {author}
  {\bibfnamefont {D.}~\bibnamefont {Kim}}, \bibinfo {author} {\bibfnamefont
  {Y.}~\bibnamefont {Lou}}, \bibinfo {author} {\bibfnamefont {C.}~\bibnamefont
  {Xu}}, \bibinfo {author} {\bibfnamefont {M.}~\bibnamefont {Li}}, \bibinfo
  {author} {\bibfnamefont {Z.}~\bibnamefont {Wei}}, \bibinfo {author}
  {\bibfnamefont {Y.}~\bibnamefont {Zhang}}, \bibinfo {author} {\bibfnamefont
  {M.~T.}\ \bibnamefont {Edmonds}}, \bibinfo {author} {\bibfnamefont
  {S.}~\bibnamefont {Li}}, \bibinfo {author} {\bibfnamefont {J.}~\bibnamefont
  {Seidel}}, \bibinfo {author} {\bibfnamefont {Y.}~\bibnamefont {Zhu}},
  \bibinfo {author} {\bibfnamefont {J.~Z.}\ \bibnamefont {Liu}}, \bibinfo
  {author} {\bibfnamefont {W.-X.}\ \bibnamefont {Tang}}, \ and\ \bibinfo
  {author} {\bibfnamefont {M.~S.}\ \bibnamefont {Fuhrer}},\ }\href {\doibase
  10.1126/sciadv.aar7720} {\bibfield  {journal} {\bibinfo  {journal} {Sci.
  Adv.}\ }\textbf {\bibinfo {volume} {4}},\ \bibinfo {pages} {8} (\bibinfo
  {year} {2018})}\BibitemShut {NoStop}%
\bibitem [{\citenamefont {Lin}\ \emph {et~al.}(2013)\citenamefont {Lin},
  \citenamefont {Wu}, \citenamefont {Zhou}, \citenamefont {Huang},
  \citenamefont {Jiang}, \citenamefont {Zheng}, \citenamefont {Zhao},
  \citenamefont {Jin}, \citenamefont {Guo}, \citenamefont {Peng},\ and\
  \citenamefont {Liu}}]{linJACS13}%
  \BibitemOpen
  \bibfield  {author} {\bibinfo {author} {\bibfnamefont {M.}~\bibnamefont
  {Lin}}, \bibinfo {author} {\bibfnamefont {D.}~\bibnamefont {Wu}}, \bibinfo
  {author} {\bibfnamefont {Y.}~\bibnamefont {Zhou}}, \bibinfo {author}
  {\bibfnamefont {W.}~\bibnamefont {Huang}}, \bibinfo {author} {\bibfnamefont
  {W.}~\bibnamefont {Jiang}}, \bibinfo {author} {\bibfnamefont
  {W.}~\bibnamefont {Zheng}}, \bibinfo {author} {\bibfnamefont
  {S.}~\bibnamefont {Zhao}}, \bibinfo {author} {\bibfnamefont {C.}~\bibnamefont
  {Jin}}, \bibinfo {author} {\bibfnamefont {Y.}~\bibnamefont {Guo}}, \bibinfo
  {author} {\bibfnamefont {H.}~\bibnamefont {Peng}}, \ and\ \bibinfo {author}
  {\bibfnamefont {Z.}~\bibnamefont {Liu}},\ }\href {\doibase 10.1021/ja406351u}
  {\bibfield  {journal} {\bibinfo  {journal} {J. Am. Chem. Soc.}\ }\textbf
  {\bibinfo {volume} {135}},\ \bibinfo {pages} {13274} (\bibinfo {year}
  {2013})}\BibitemShut {NoStop}%
\bibitem [{\citenamefont {Zheng}\ \emph {et~al.}(2015)\citenamefont {Zheng},
  \citenamefont {Xie}, \citenamefont {Zhou}, \citenamefont {Chen},
  \citenamefont {Jiang}, \citenamefont {Zhao}, \citenamefont {Wu},
  \citenamefont {Jing}, \citenamefont {Wu}, \citenamefont {Chen}, \citenamefont
  {Guo}, \citenamefont {Yin}, \citenamefont {Huang}, \citenamefont {Xu},
  \citenamefont {Liu},\ and\ \citenamefont {Peng}}]{zhengNC15}%
  \BibitemOpen
  \bibfield  {author} {\bibinfo {author} {\bibfnamefont {W.}~\bibnamefont
  {Zheng}}, \bibinfo {author} {\bibfnamefont {T.}~\bibnamefont {Xie}}, \bibinfo
  {author} {\bibfnamefont {Y.}~\bibnamefont {Zhou}}, \bibinfo {author}
  {\bibfnamefont {Y.~L.}\ \bibnamefont {Chen}}, \bibinfo {author}
  {\bibfnamefont {W.}~\bibnamefont {Jiang}}, \bibinfo {author} {\bibfnamefont
  {S.}~\bibnamefont {Zhao}}, \bibinfo {author} {\bibfnamefont {J.}~\bibnamefont
  {Wu}}, \bibinfo {author} {\bibfnamefont {Y.}~\bibnamefont {Jing}}, \bibinfo
  {author} {\bibfnamefont {Y.}~\bibnamefont {Wu}}, \bibinfo {author}
  {\bibfnamefont {G.}~\bibnamefont {Chen}}, \bibinfo {author} {\bibfnamefont
  {Y.}~\bibnamefont {Guo}}, \bibinfo {author} {\bibfnamefont {J.}~\bibnamefont
  {Yin}}, \bibinfo {author} {\bibfnamefont {S.}~\bibnamefont {Huang}}, \bibinfo
  {author} {\bibfnamefont {H.~Q.}\ \bibnamefont {Xu}}, \bibinfo {author}
  {\bibfnamefont {Z.}~\bibnamefont {Liu}}, \ and\ \bibinfo {author}
  {\bibfnamefont {H.}~\bibnamefont {Peng}},\ }\href {\doibase
  10.1038/ncomms7972} {\bibfield  {journal} {\bibinfo  {journal} {Nat.
  Commun.}\ }\textbf {\bibinfo {volume} {6}},\ \bibinfo {pages} {6972}
  (\bibinfo {year} {2015})}\BibitemShut {NoStop}%
\bibitem [{\citenamefont {Giannozzi}\ \emph {et~al.}(2017)\citenamefont
  {Giannozzi}, \citenamefont {Andreussi}, \citenamefont {Brumme}, \citenamefont
  {Bunau}, \citenamefont {Buongiorno~Nardelli}, \citenamefont {Calandra},
  \citenamefont {Car}, \citenamefont {Cavazzoni}, \citenamefont {Ceresoli},
  \citenamefont {Cococcioni}, \citenamefont {Colonna}, \citenamefont
  {Carnimeo}, \citenamefont {Dal~Corso}, \citenamefont {de~Gironcoli},
  \citenamefont {Delugas}, \citenamefont {DiStasio}, \citenamefont {Ferretti},
  \citenamefont {Floris}, \citenamefont {Fratesi}, \citenamefont {Fugallo},
  \citenamefont {Gebauer}, \citenamefont {Gerstmann}, \citenamefont {Giustino},
  \citenamefont {Gorni}, \citenamefont {Jia}, \citenamefont {Kawamura},
  \citenamefont {Ko}, \citenamefont {Kokalj}, \citenamefont
  {K{\"u}{\c{c}}{\"u}kbenli}, \citenamefont {Lazzeri}, \citenamefont {Marsili},
  \citenamefont {Marzari}, \citenamefont {Mauri}, \citenamefont {Nguyen},
  \citenamefont {Nguyen}, \citenamefont {Otero-de-la Roza}, \citenamefont
  {Paulatto}, \citenamefont {Ponc{\'e}}, \citenamefont {Rocca}, \citenamefont
  {Sabatini}, \citenamefont {Santra}, \citenamefont {Schlipf}, \citenamefont
  {Seitsonen}, \citenamefont {Smogunov}, \citenamefont {Timrov}, \citenamefont
  {Thonhauser}, \citenamefont {Umari}, \citenamefont {Vast}, \citenamefont
  {Wu},\ and\ \citenamefont {Baroni}}]{giannozziJPCM17}%
  \BibitemOpen
  \bibfield  {author} {\bibinfo {author} {\bibfnamefont {P.}~\bibnamefont
  {Giannozzi}}, \bibinfo {author} {\bibfnamefont {O.}~\bibnamefont
  {Andreussi}}, \bibinfo {author} {\bibfnamefont {T.}~\bibnamefont {Brumme}},
  \bibinfo {author} {\bibfnamefont {O.}~\bibnamefont {Bunau}}, \bibinfo
  {author} {\bibfnamefont {M.}~\bibnamefont {Buongiorno~Nardelli}}, \bibinfo
  {author} {\bibfnamefont {M.}~\bibnamefont {Calandra}}, \bibinfo {author}
  {\bibfnamefont {R.}~\bibnamefont {Car}}, \bibinfo {author} {\bibfnamefont
  {C.}~\bibnamefont {Cavazzoni}}, \bibinfo {author} {\bibfnamefont
  {D.}~\bibnamefont {Ceresoli}}, \bibinfo {author} {\bibfnamefont
  {M.}~\bibnamefont {Cococcioni}}, \bibinfo {author} {\bibfnamefont
  {N.}~\bibnamefont {Colonna}}, \bibinfo {author} {\bibfnamefont
  {I.}~\bibnamefont {Carnimeo}}, \bibinfo {author} {\bibfnamefont
  {A.}~\bibnamefont {Dal~Corso}}, \bibinfo {author} {\bibfnamefont
  {S.}~\bibnamefont {de~Gironcoli}}, \bibinfo {author} {\bibfnamefont
  {P.}~\bibnamefont {Delugas}}, \bibinfo {author} {\bibfnamefont {R.~A.}\
  \bibnamefont {DiStasio}}, \bibinfo {author} {\bibfnamefont {A.}~\bibnamefont
  {Ferretti}}, \bibinfo {author} {\bibfnamefont {A.}~\bibnamefont {Floris}},
  \bibinfo {author} {\bibfnamefont {G.}~\bibnamefont {Fratesi}}, \bibinfo
  {author} {\bibfnamefont {G.}~\bibnamefont {Fugallo}}, \bibinfo {author}
  {\bibfnamefont {R.}~\bibnamefont {Gebauer}}, \bibinfo {author} {\bibfnamefont
  {U.}~\bibnamefont {Gerstmann}}, \bibinfo {author} {\bibfnamefont
  {F.}~\bibnamefont {Giustino}}, \bibinfo {author} {\bibfnamefont
  {T.}~\bibnamefont {Gorni}}, \bibinfo {author} {\bibfnamefont
  {J.}~\bibnamefont {Jia}}, \bibinfo {author} {\bibfnamefont {M.}~\bibnamefont
  {Kawamura}}, \bibinfo {author} {\bibfnamefont {H.-Y.}\ \bibnamefont {Ko}},
  \bibinfo {author} {\bibfnamefont {A.}~\bibnamefont {Kokalj}}, \bibinfo
  {author} {\bibfnamefont {E.}~\bibnamefont {K{\"u}{\c{c}}{\"u}kbenli}},
  \bibinfo {author} {\bibfnamefont {M.}~\bibnamefont {Lazzeri}}, \bibinfo
  {author} {\bibfnamefont {M.}~\bibnamefont {Marsili}}, \bibinfo {author}
  {\bibfnamefont {N.}~\bibnamefont {Marzari}}, \bibinfo {author} {\bibfnamefont
  {F.}~\bibnamefont {Mauri}}, \bibinfo {author} {\bibfnamefont {N.~L.}\
  \bibnamefont {Nguyen}}, \bibinfo {author} {\bibfnamefont {H.-V.}\
  \bibnamefont {Nguyen}}, \bibinfo {author} {\bibfnamefont {A.}~\bibnamefont
  {Otero-de-la Roza}}, \bibinfo {author} {\bibfnamefont {L.}~\bibnamefont
  {Paulatto}}, \bibinfo {author} {\bibfnamefont {S.}~\bibnamefont {Ponc{\'e}}},
  \bibinfo {author} {\bibfnamefont {D.}~\bibnamefont {Rocca}}, \bibinfo
  {author} {\bibfnamefont {R.}~\bibnamefont {Sabatini}}, \bibinfo {author}
  {\bibfnamefont {B.}~\bibnamefont {Santra}}, \bibinfo {author} {\bibfnamefont
  {M.}~\bibnamefont {Schlipf}}, \bibinfo {author} {\bibfnamefont {A.~P.}\
  \bibnamefont {Seitsonen}}, \bibinfo {author} {\bibfnamefont {A.}~\bibnamefont
  {Smogunov}}, \bibinfo {author} {\bibfnamefont {I.}~\bibnamefont {Timrov}},
  \bibinfo {author} {\bibfnamefont {T.}~\bibnamefont {Thonhauser}}, \bibinfo
  {author} {\bibfnamefont {P.}~\bibnamefont {Umari}}, \bibinfo {author}
  {\bibfnamefont {N.}~\bibnamefont {Vast}}, \bibinfo {author} {\bibfnamefont
  {X.}~\bibnamefont {Wu}}, \ and\ \bibinfo {author} {\bibfnamefont
  {S.}~\bibnamefont {Baroni}},\ }\href {\doibase 10.1088/1361-648X/aa8f79}
  {\bibfield  {journal} {\bibinfo  {journal} {J. Phys. Condens. Matter}\
  }\textbf {\bibinfo {volume} {29}},\ \bibinfo {pages} {465901} (\bibinfo
  {year} {2017})}\BibitemShut {NoStop}%
\bibitem [{\citenamefont {Mendoza}\ \emph {et~al.}()\citenamefont {Mendoza},
  \citenamefont {Anderson}, \citenamefont {Cabellos},\ and\ \citenamefont
  {Rangel}}]{tiniba}%
  \BibitemOpen
  \bibfield  {author} {\bibinfo {author} {\bibfnamefont {B.~S.}\ \bibnamefont
  {Mendoza}}, \bibinfo {author} {\bibfnamefont {S.~M.}\ \bibnamefont
  {Anderson}}, \bibinfo {author} {\bibfnamefont {J.~L.}\ \bibnamefont
  {Cabellos}}, \ and\ \bibinfo {author} {\bibfnamefont {T.}~\bibnamefont
  {Rangel}},\ }\href@noop {} {\enquote {\bibinfo {title} {{TINIBA}: \emph{Ab
  initio} calculation of the optical properties of solids, surfaces,
  interfaces, and {2D} materials},}\ }\bibinfo {note} {INDAUTOR-Mexico No.
  03-2009-120114033400-01}\BibitemShut {NoStop}%
\bibitem [{\citenamefont {Anderson}\ \emph {et~al.}(2015)\citenamefont
  {Anderson}, \citenamefont {Tancogne-Dejean}, \citenamefont {Mendoza},\ and\
  \citenamefont {V{\'e}niard}}]{andersonPRB15}%
  \BibitemOpen
  \bibfield  {author} {\bibinfo {author} {\bibfnamefont {S.~M.}\ \bibnamefont
  {Anderson}}, \bibinfo {author} {\bibfnamefont {N.}~\bibnamefont
  {Tancogne-Dejean}}, \bibinfo {author} {\bibfnamefont {B.~S.}\ \bibnamefont
  {Mendoza}}, \ and\ \bibinfo {author} {\bibfnamefont {V.}~\bibnamefont
  {V{\'e}niard}},\ }\href {\doibase 10.1103/PhysRevB.91.075302} {\bibfield
  {journal} {\bibinfo  {journal} {Phys. Rev. B}\ }\textbf {\bibinfo {volume}
  {91}},\ \bibinfo {pages} {075302} (\bibinfo {year} {2015})}\BibitemShut
  {NoStop}%
\bibitem [{\citenamefont {Gonze}\ \emph {et~al.}(2009)\citenamefont {Gonze},
  \citenamefont {Amadon}, \citenamefont {Anglade}, \citenamefont {Beuken},
  \citenamefont {Bottin}, \citenamefont {Boulanger}, \citenamefont {Bruneval},
  \citenamefont {Caliste}, \citenamefont {Caracas}, \citenamefont
  {C{\^{o}}t{\'{e}}}, \citenamefont {Deutsch}, \citenamefont {Genovese},
  \citenamefont {Ghosez}, \citenamefont {Giantomassi}, \citenamefont
  {Goedecker}, \citenamefont {Hamann}, \citenamefont {Hermet}, \citenamefont
  {Jollet}, \citenamefont {Jomard}, \citenamefont {Leroux}, \citenamefont
  {Mancini}, \citenamefont {Mazevet}, \citenamefont {Oliveira}, \citenamefont
  {Onida}, \citenamefont {Pouillon}, \citenamefont {Rangel}, \citenamefont
  {Rignanese}, \citenamefont {Sangalli}, \citenamefont {Shaltaf}, \citenamefont
  {Torrent}, \citenamefont {Verstraete}, \citenamefont {Zerah},\ and\
  \citenamefont {Zwanziger}}]{gonzeCPS09}%
  \BibitemOpen
  \bibfield  {author} {\bibinfo {author} {\bibfnamefont {X.}~\bibnamefont
  {Gonze}}, \bibinfo {author} {\bibfnamefont {B.}~\bibnamefont {Amadon}},
  \bibinfo {author} {\bibfnamefont {P.-M.}\ \bibnamefont {Anglade}}, \bibinfo
  {author} {\bibfnamefont {J.-M.}\ \bibnamefont {Beuken}}, \bibinfo {author}
  {\bibfnamefont {F.}~\bibnamefont {Bottin}}, \bibinfo {author} {\bibfnamefont
  {P.}~\bibnamefont {Boulanger}}, \bibinfo {author} {\bibfnamefont
  {F.}~\bibnamefont {Bruneval}}, \bibinfo {author} {\bibfnamefont
  {D.}~\bibnamefont {Caliste}}, \bibinfo {author} {\bibfnamefont
  {R.}~\bibnamefont {Caracas}}, \bibinfo {author} {\bibfnamefont
  {M.}~\bibnamefont {C{\^{o}}t{\'{e}}}}, \bibinfo {author} {\bibfnamefont
  {T.}~\bibnamefont {Deutsch}}, \bibinfo {author} {\bibfnamefont
  {L.}~\bibnamefont {Genovese}}, \bibinfo {author} {\bibfnamefont
  {P.}~\bibnamefont {Ghosez}}, \bibinfo {author} {\bibfnamefont
  {M.}~\bibnamefont {Giantomassi}}, \bibinfo {author} {\bibfnamefont
  {S.}~\bibnamefont {Goedecker}}, \bibinfo {author} {\bibfnamefont {D.~R.}\
  \bibnamefont {Hamann}}, \bibinfo {author} {\bibfnamefont {P.}~\bibnamefont
  {Hermet}}, \bibinfo {author} {\bibfnamefont {F.}~\bibnamefont {Jollet}},
  \bibinfo {author} {\bibfnamefont {G.}~\bibnamefont {Jomard}}, \bibinfo
  {author} {\bibfnamefont {S.}~\bibnamefont {Leroux}}, \bibinfo {author}
  {\bibfnamefont {M.}~\bibnamefont {Mancini}}, \bibinfo {author} {\bibfnamefont
  {S.}~\bibnamefont {Mazevet}}, \bibinfo {author} {\bibfnamefont {M.~J.~T.}\
  \bibnamefont {Oliveira}}, \bibinfo {author} {\bibfnamefont {G.}~\bibnamefont
  {Onida}}, \bibinfo {author} {\bibfnamefont {Y.}~\bibnamefont {Pouillon}},
  \bibinfo {author} {\bibfnamefont {T.}~\bibnamefont {Rangel}}, \bibinfo
  {author} {\bibfnamefont {G.-M.}\ \bibnamefont {Rignanese}}, \bibinfo {author}
  {\bibfnamefont {D.}~\bibnamefont {Sangalli}}, \bibinfo {author}
  {\bibfnamefont {R.}~\bibnamefont {Shaltaf}}, \bibinfo {author} {\bibfnamefont
  {M.}~\bibnamefont {Torrent}}, \bibinfo {author} {\bibfnamefont {M.~J.}\
  \bibnamefont {Verstraete}}, \bibinfo {author} {\bibfnamefont
  {G.}~\bibnamefont {Zerah}}, \ and\ \bibinfo {author} {\bibfnamefont {J.~W.}\
  \bibnamefont {Zwanziger}},\ }\href {\doibase 10.1016/j.cpc.2009.07.007}
  {\bibfield  {journal} {\bibinfo  {journal} {Comp. Phys. Commun.}\ }\textbf
  {\bibinfo {volume} {180}},\ \bibinfo {pages} {2582} (\bibinfo {year}
  {2009})}\BibitemShut {NoStop}%
\bibitem [{abi()}]{abinit}%
  \BibitemOpen
  \href@noop {} {}\bibinfo {note} {The ABINIT code is a common project of the
  Universit{\'e} Catholique de Louvain, Corning Incorporated, and other
  contributors (URL http://www.abinit.org).}\BibitemShut {Stop}%
\bibitem [{\citenamefont {Troullier}\ and\ \citenamefont
  {Martins}(1991)}]{troullierPRB91}%
  \BibitemOpen
  \bibfield  {author} {\bibinfo {author} {\bibfnamefont {N.}~\bibnamefont
  {Troullier}}\ and\ \bibinfo {author} {\bibfnamefont {J.~L.}\ \bibnamefont
  {Martins}},\ }\href {\doibase 10.1103/PhysRevB.43.1993} {\bibfield  {journal}
  {\bibinfo  {journal} {Phys. Rev. B}\ }\textbf {\bibinfo {volume} {43}},\
  \bibinfo {pages} {1993} (\bibinfo {year} {1991})}\BibitemShut {NoStop}%
\bibitem [{\citenamefont {Olevano}\ \emph {et~al.}()\citenamefont {Olevano},
  \citenamefont {Reining},\ and\ \citenamefont {Sottile}}]{olevanoDP}%
  \BibitemOpen
  \bibfield  {author} {\bibinfo {author} {\bibfnamefont {V.}~\bibnamefont
  {Olevano}}, \bibinfo {author} {\bibfnamefont {L.}~\bibnamefont {Reining}}, \
  and\ \bibinfo {author} {\bibfnamefont {F.}~\bibnamefont {Sottile}},\
  }\href@noop {} {}\bibinfo {note} {\url{http://dp-code.org}}\BibitemShut
  {NoStop}%
\bibitem [{\citenamefont {Reining}\ \emph {et~al.}()\citenamefont {Reining},
  \citenamefont {Olevano}, \citenamefont {Sottile}, \citenamefont {Albrecht},\
  and\ \citenamefont {Onida}}]{reiningEXC}%
  \BibitemOpen
  \bibfield  {author} {\bibinfo {author} {\bibfnamefont {L.}~\bibnamefont
  {Reining}}, \bibinfo {author} {\bibfnamefont {V.}~\bibnamefont {Olevano}},
  \bibinfo {author} {\bibfnamefont {F.}~\bibnamefont {Sottile}}, \bibinfo
  {author} {\bibfnamefont {S.}~\bibnamefont {Albrecht}}, \ and\ \bibinfo
  {author} {\bibfnamefont {G.}~\bibnamefont {Onida}},\ }\href@noop {}
  {}\bibinfo {howpublished} {\url{http://www.bethe-salpeter.org}}\BibitemShut
  {NoStop}%
\bibitem [{\citenamefont {Tancogne-Dejean}\ \emph {et~al.}(2015)\citenamefont
  {Tancogne-Dejean}, \citenamefont {Giorgetti},\ and\ \citenamefont
  {V{\'e}niard}}]{tancognePRB15}%
  \BibitemOpen
  \bibfield  {author} {\bibinfo {author} {\bibfnamefont {N.}~\bibnamefont
  {Tancogne-Dejean}}, \bibinfo {author} {\bibfnamefont {C.}~\bibnamefont
  {Giorgetti}}, \ and\ \bibinfo {author} {\bibfnamefont {V.}~\bibnamefont
  {V{\'e}niard}},\ }\href {\doibase 10.1103/PhysRevB.92.245308} {\bibfield
  {journal} {\bibinfo  {journal} {Phys. Rev. B}\ }\textbf {\bibinfo {volume}
  {92}},\ \bibinfo {pages} {245308} (\bibinfo {year} {2015})}\BibitemShut
  {NoStop}%
\bibitem [{\citenamefont {Hedin}(1965)}]{hedinPR65}%
  \BibitemOpen
  \bibfield  {author} {\bibinfo {author} {\bibfnamefont {L.}~\bibnamefont
  {Hedin}},\ }\href {http://link.aps.org/doi/10.1103/PhysRev.139.A796}
  {\bibfield  {journal} {\bibinfo  {journal} {Phys. Rev.}\ }\textbf {\bibinfo
  {volume} {139}},\ \bibinfo {pages} {A796} (\bibinfo {year}
  {1965})}\BibitemShut {NoStop}%
\bibitem [{\citenamefont {Boyd}(2003)}]{boyd}%
  \BibitemOpen
  \bibfield  {author} {\bibinfo {author} {\bibfnamefont {R.}~\bibnamefont
  {Boyd}},\ }\href@noop {} {\emph {\bibinfo {title} {Nonlinear Optics}}}\
  (\bibinfo  {publisher} {Academic Press},\ \bibinfo {address} {New York},\
  \bibinfo {year} {2003})\BibitemShut {NoStop}%
\bibitem [{\citenamefont {Popov}\ \emph {et~al.}(1995)\citenamefont {Popov},
  \citenamefont {Svirko},\ and\ \citenamefont {Zheludev}}]{popovbook}%
  \BibitemOpen
  \bibfield  {author} {\bibinfo {author} {\bibfnamefont {S.~V.}\ \bibnamefont
  {Popov}}, \bibinfo {author} {\bibfnamefont {Y.~P.}\ \bibnamefont {Svirko}}, \
  and\ \bibinfo {author} {\bibfnamefont {N.~I.}\ \bibnamefont {Zheludev}},\
  }\href@noop {} {\emph {\bibinfo {title} {Susceptibility tensors for nonlinear
  optics}}}\ (\bibinfo  {publisher} {CRC Press},\ \bibinfo {year}
  {1995})\BibitemShut {NoStop}%
\bibitem [{\citenamefont {Sipe}\ \emph {et~al.}(1987)\citenamefont {Sipe},
  \citenamefont {Moss},\ and\ \citenamefont {van Driel}}]{sipePRB87}%
  \BibitemOpen
  \bibfield  {author} {\bibinfo {author} {\bibfnamefont {J.~E.}\ \bibnamefont
  {Sipe}}, \bibinfo {author} {\bibfnamefont {D.~J.}\ \bibnamefont {Moss}}, \
  and\ \bibinfo {author} {\bibfnamefont {H.~M.}\ \bibnamefont {van Driel}},\
  }\href {\doibase 10.1103/PhysRevB.35.1129} {\bibfield  {journal} {\bibinfo
  {journal} {Phys. Rev. B}\ }\textbf {\bibinfo {volume} {35}},\ \bibinfo
  {pages} {1129} (\bibinfo {year} {1987})}\BibitemShut {NoStop}%
\bibitem [{\citenamefont {Yamada}\ and\ \citenamefont
  {Kimura}(1994)}]{yamadaPRB94}%
  \BibitemOpen
  \bibfield  {author} {\bibinfo {author} {\bibfnamefont {C.}~\bibnamefont
  {Yamada}}\ and\ \bibinfo {author} {\bibfnamefont {T.}~\bibnamefont
  {Kimura}},\ }\href {\doibase 10.1103/PhysRevB.49.14372} {\bibfield  {journal}
  {\bibinfo  {journal} {Phys. Rev. B}\ }\textbf {\bibinfo {volume} {49}},\
  \bibinfo {pages} {14372} (\bibinfo {year} {1994})}\BibitemShut {NoStop}%
\bibitem [{\citenamefont {Plechinger}\ \emph {et~al.}(2015)\citenamefont
  {Plechinger}, \citenamefont {Mooshammer}, \citenamefont {Castellanos-Gomez},
  \citenamefont {Steele}, \citenamefont {Sch\"{u}ller},\ and\ \citenamefont
  {Korn}}]{plechinger2DM15}%
  \BibitemOpen
  \bibfield  {author} {\bibinfo {author} {\bibfnamefont {G.}~\bibnamefont
  {Plechinger}}, \bibinfo {author} {\bibfnamefont {F.}~\bibnamefont
  {Mooshammer}}, \bibinfo {author} {\bibfnamefont {A.}~\bibnamefont
  {Castellanos-Gomez}}, \bibinfo {author} {\bibfnamefont {G.~A.}\ \bibnamefont
  {Steele}}, \bibinfo {author} {\bibfnamefont {C.}~\bibnamefont
  {Sch\"{u}ller}}, \ and\ \bibinfo {author} {\bibfnamefont {T.}~\bibnamefont
  {Korn}},\ }\href {\doibase 10.1088/2053-1583/2/3/034016} {\bibfield
  {journal} {\bibinfo  {journal} {2D Materials}\ }\textbf {\bibinfo {volume}
  {2}},\ \bibinfo {pages} {034016} (\bibinfo {year} {2015})}\BibitemShut
  {NoStop}%
\bibitem [{\citenamefont {Debbichi}\ \emph {et~al.}(2015)\citenamefont
  {Debbichi}, \citenamefont {Eriksson},\ and\ \citenamefont
  {Leb{\'e}gue}}]{debbichiJPCL15}%
  \BibitemOpen
  \bibfield  {author} {\bibinfo {author} {\bibfnamefont {L.}~\bibnamefont
  {Debbichi}}, \bibinfo {author} {\bibfnamefont {O.}~\bibnamefont {Eriksson}},
  \ and\ \bibinfo {author} {\bibfnamefont {S.}~\bibnamefont {Leb{\'e}gue}},\
  }\href {\doibase 10.1021/acs.jpclett.5b01356} {\bibfield  {journal} {\bibinfo
   {journal} {J. Phys. Chem. Lett.}\ }\textbf {\bibinfo {volume} {6}},\
  \bibinfo {pages} {3098} (\bibinfo {year} {2015})}\BibitemShut {NoStop}%
\bibitem [{\citenamefont {Shkrebtii}\ \emph {et~al.}(2019)\citenamefont
  {Shkrebtii}, \citenamefont {Minnings}, \citenamefont {Perinparajah},
  \citenamefont {Arzate}, \citenamefont {Anderson}, \citenamefont {Mendoza},
  \citenamefont {Cho}, \citenamefont {Downer},\ and\ \citenamefont
  {Zahn}}]{shkrebtiiproc}%
  \BibitemOpen
  \bibfield  {author} {\bibinfo {author} {\bibfnamefont {A.~I.}\ \bibnamefont
  {Shkrebtii}}, \bibinfo {author} {\bibfnamefont {R.}~\bibnamefont {Minnings}},
  \bibinfo {author} {\bibfnamefont {G.}~\bibnamefont {Perinparajah}}, \bibinfo
  {author} {\bibfnamefont {N.}~\bibnamefont {Arzate}}, \bibinfo {author}
  {\bibfnamefont {S.}~\bibnamefont {Anderson}}, \bibinfo {author}
  {\bibfnamefont {B.}~\bibnamefont {Mendoza}}, \bibinfo {author} {\bibfnamefont
  {Y.}~\bibnamefont {Cho}}, \bibinfo {author} {\bibfnamefont {M.}~\bibnamefont
  {Downer}}, \ and\ \bibinfo {author} {\bibfnamefont {D.}~\bibnamefont
  {Zahn}},\ }in\ \href@noop {} {\emph {\bibinfo {booktitle} {ANM 2019: Advanced
  Nano Materials Conference}}}\ (\bibinfo {address} {Aveiro, Portugal},\
  \bibinfo {year} {2019})\ \bibinfo {note} {to be submitted to proceedings on
  Sep 30.}\BibitemShut {Stop}%
\bibitem [{\citenamefont {Gusakova}\ \emph {et~al.}(2017)\citenamefont
  {Gusakova}, \citenamefont {Wang}, \citenamefont {Shiau}, \citenamefont
  {Krivosheeva}, \citenamefont {Shaposhnikov}, \citenamefont {Borisenko},
  \citenamefont {Gusakov},\ and\ \citenamefont {Tay}}]{gusakovaPSSA17}%
  \BibitemOpen
  \bibfield  {author} {\bibinfo {author} {\bibfnamefont {J.}~\bibnamefont
  {Gusakova}}, \bibinfo {author} {\bibfnamefont {X.}~\bibnamefont {Wang}},
  \bibinfo {author} {\bibfnamefont {L.~L.}\ \bibnamefont {Shiau}}, \bibinfo
  {author} {\bibfnamefont {A.}~\bibnamefont {Krivosheeva}}, \bibinfo {author}
  {\bibfnamefont {V.}~\bibnamefont {Shaposhnikov}}, \bibinfo {author}
  {\bibfnamefont {V.}~\bibnamefont {Borisenko}}, \bibinfo {author}
  {\bibfnamefont {V.}~\bibnamefont {Gusakov}}, \ and\ \bibinfo {author}
  {\bibfnamefont {B.~K.}\ \bibnamefont {Tay}},\ }\href {\doibase
  10.1002/pssa.201700218} {\bibfield  {journal} {\bibinfo  {journal} {Phys.
  Status Solidi A}\ }\textbf {\bibinfo {volume} {214}},\ \bibinfo {pages}
  {1700218} (\bibinfo {year} {2017})}\BibitemShut {NoStop}%
\bibitem [{\citenamefont {Zheng}\ \emph {et~al.}(2016)\citenamefont {Zheng},
  \citenamefont {Yao},\ and\ \citenamefont {Yang}}]{zhengJMCC16}%
  \BibitemOpen
  \bibfield  {author} {\bibinfo {author} {\bibfnamefont {Z.~Q.}\ \bibnamefont
  {Zheng}}, \bibinfo {author} {\bibfnamefont {J.~D.}\ \bibnamefont {Yao}}, \
  and\ \bibinfo {author} {\bibfnamefont {G.~W.}\ \bibnamefont {Yang}},\ }\href
  {\doibase 10.1039/C6TC02296K} {\bibfield  {journal} {\bibinfo  {journal} {J.
  Mater. Chem. C}\ }\textbf {\bibinfo {volume} {4}},\ \bibinfo {pages} {8094}
  (\bibinfo {year} {2016})}\BibitemShut {NoStop}%
\end{thebibliography}

\end{document}